%% file: SiAnneal.tex
\documentclass[preprint,12pt]{elsarticle}

\usepackage{lineno}

\usepackage{graphicx}
\usepackage{amsmath}
\usepackage{amssymb}
\usepackage{xspace}
\usepackage{natbib}
\usepackage{dcolumn}
\usepackage{tabularx}
\usepackage{url}

\newcommand{\Mrad}{\ensuremath{\mathrm{Mrad}}\xspace}

\newcommand{\degC}{\ensuremath{{}^{\circ}\mathrm{C}}\xspace}
\newcommand{\iv}{$IV$\xspace}
\newcommand{\ibias}{\ensuremath{I_\mathrm{bias}}\xspace}

\newcommand{\vdep}{\ensuremath{V_\mathrm{dep}}\xspace}
\newcommand{\vdepmap}{\ensuremath{\widetilde{V}_\mathrm{dep}}\xspace}
\newcommand{\vknee}{\ensuremath{V_\mathrm{knee}}\xspace}
\newcommand{\vbias}{\ensuremath{V_\mathrm{bias}}\xspace}
\newcommand{\vop}{\ensuremath{V_\mathrm{op}}\xspace}
\newcommand{\uA}{\ensuremath{\mu\mathrm{A}}\xspace}

\newcommand{\rphi}{$r$-\textit{$\phi$}}
\newcommand{\rz}{$r$-$z$}

\newcommand{\degreesC}{\degC}

\journal{Nuclear Instrumentation Methods A}

\begin{document}
\begin{frontmatter}

%
%
\title{CDF Run II Silicon Vertex Detector Annealing Study}

%
%
\input{author_list.tex}

%
%
\input{abstract.tex}

%
%
\begin{keyword}
Annealing \sep Silicon \sep Vertex Detector \sep
CDF \sep Detector Operations 


\end{keyword}
\end{frontmatter}

%
%
\newpage
\tableofcontents
%

\newpage

%
%

\input{introduction.tex}

\input{anneal_theory.tex}

\input{detector.tex}

\input{procedure.tex}

\input{currents.tex}

\input{dv.tex}

\input{results.tex}

\input{summary.tex}

%
%
\appendix
\input{sh.tex}

%
%
\input{SiAnneal_bib}

\end{document}

%% file: author_list.tex
%
%
%


\author[FNAL]{M.~Stancari}
\author[FNAL]{K.~Knoepfel}
\author[FNAL]{S.~Behari}
\author[FNAL]{D.~Christian}
\author[FNAL,BNL]{B.~Di~Ruzza}
\author[FNAL]{S.~Jindariani}
\author[FNAL]{T.~R.~Junk}
\author[WAYNE]{M.~Mattson}
\author[SINICA] {A.~Mitra}
\author[FNAL]{M.~N.~Mondragon}
\author[FNAL]{A.~Sukhanov}

\address[SINICA]{Academia Sinica, Taipei, Taiwan 11529, Republic of China}
\address[BNL]{Physics Department, Brookhaven National Laboratory, Upton, New York 11973}
\address[FNAL]{Fermi National Accelerator Laboratory, Batavia, Illinois 60510}
\address[WAYNE]{Wayne State University, Detroit, Michigan 48202}

%% file: abstract.tex
%
%
\begin{abstract}
  Between Run II commissioning in early 2001 and the end of operations
  in September 2011, the Tevatron collider delivered 12~fb$^{-1}$ of
  $p\bar{p}$ collisions at $\sqrt{s}=1.96$~TeV to the Collider
  Detector at Fermilab (CDF).  During that time, the CDF silicon
  vertex detector was subject to radiation doses of up to 12~Mrad.
  After the end of operations, the silicon detector was annealed for
  24 days at 18\degC.  In this paper, we present a measurement of the
  change in the bias currents for a subset of sensors during the
  annealing period.  We also introduce a novel method for monitoring
  the depletion voltage throughout the annealing period.  The observed
  bias current evolution can be characterized by a falling exponential
  term with time constant 
  $\tau_I=17.88\pm0.36$(stat.)$\pm0.25$(syst.)~days.
  We observe an average decrease of $(27\pm3)\%$ in the
  depletion voltage, whose evolution can similarly be described by an
  exponential time constant of $\tau_V=6.21\pm0.21$~days.  These
  results are consistent with the Hamburg model 
within the measurement uncertainties.

\end{abstract}

%% file: introduction.tex
%
%
\section{Introduction}
\label{sec:intro}

In high-energy physics (HEP) experiments, silicon sensors serve a
crucial role in the detection of charged particle positions, momenta,
and to some extent, $dE/dX$ information.  Due to their typically close
location to the collision point of hadron colliders, silicon sensors
often incur intense radiation damage due to the numerous particles
from collisions that traverse them.  The macroscopic effects of
radiation damage on silicon sensors in HEP detectors has been
extensively studied.  The leakage currents increase linearly with
radiation dose, and for $n$-bulk sensors, the depletion voltage \vdep
initially decreases until the sensor appears to undergo type
inversion, at which point \vdep then increases with radiation dose.
These macroscopic changes have been linked to the formation of crystal
defects when atoms are displaced from their lattice positions after
collisions with particles from the radiation field.

The process of annealing is the opposite effect, where increasing the
temperature of the silicon sensor allows the displaced atoms to settle
back into vacant lattice positions, eliminating the local crystal
defect, and at least partially recovering some of the initial behavior
of the silicon sensor as it was before irradiation.  Annealing, which
is strongly temperature dependent, has been studied with test sensors,
where the irradiation phase and the annealing phase can be isolated
from each other by strict temperature controls.  Such temperature
control enables the construction of silicon-behavior models which can
closely approximate ideal silicon sensor behavior.  The most popular
of them is the Hamburg
  model~\cite{Moll:1999kv,Wunstorf:1992ua} whose verification and that
other models is ongoing.

Because annealing can prolong the life of a HEP silicon detector,
understanding how the macroscopic quantities such as leakage current
and depletion voltage change with time for different temperatures is
of great interest to the HEP silicon detector community.  Test bench studies are
usually done at warm (40-80\degC) temperatures to maximize the
annealing effect in the available time, while annealing of HEP
detectors is more practical at room temperature.

This article describes the annealing studies that were performed with
the silicon detector system at the CDF experiment at Fermi National
Accelerator Laboratory.  The silicon sensors were exposed to 0.4-12
\Mrad of radiation over the course of 10 years, and dedicated
annealing studies were performed after the end of Tevatron Run
II. This \textit{in-situ} measurement of annealing with an operating
HEP detector required a new method for monitoring the depletion
voltage of the sensors.  We discuss some annealing theory in
Sec.~\ref{sec:si_an} and the detector in Sec.~\ref{sec:det}.  The
measurement and monitoring procedures are detailed in
Sec.~\ref{sec:meas}. The analysis of
the current changes and depletion voltages are given in
Secs.~\ref{sec:curr} and~\ref{sec:dv} and the results and conclusions
follow in the remaining sections.


%% file: anneal_theory.tex
\section{Silicon Annealing}
\label{sec:si_an}

The behavior of a silicon detector can be characterized by several
quantities.  For this study, we consider the leakage current, and the
depletion voltage \vdep , which for an unirradiated sensor is defined
as the minimum bias voltage applied to the sensor that can deplete it
of free charge carriers.  As the silicon sensors are irradiated, the
behavior of these quantities changes.  The leakage current increases
in a manner linearly proportional to the fluence:
\begin{linenomath}
\begin{equation}
\Delta I = \alpha \Phi_{eq}V
\end{equation}
\end{linenomath}
where $\alpha$ is the current related damage rate, $\Phi_{eq}$ is the
fluence, and $V$ is the volume of the sensor. $\Delta I$ is the
increase in leakage current from its original value $I_0$.  The
magnitude of $\alpha$ is temperature-dependent and on the order of
$10^{-17}$~A/cm.

\begin{table}[b]
  \begin{tabular*}{\linewidth}{@{\extracolsep{\fill}}l|cccccc} \hline\hline
    Term $i$                    &  1                 & 2                   & 3      & 4      & 5      & 6 \\ \hline
    $\tau_i$ at 18~\degC (days) & $1.68\times10^{-2}$ & $1.12\times10^{-1}$  & $1.02$ & $13.9$ & $83.7$ & $\infty$ \\
    $\tau_i$ at 11~\degC (days) & $4.99\times10^{-2}$ & $3.32\times10^{-1}$  & $3.05$ & $41.4$ & $249$ & $\infty$ \\
    $\tau_i$ at -5~\degC (days) & $7.46\times10^{-1}$ & $4.97$              & $45.5$ & $619$  & $3\ 720$ & $\infty$ \\
    \hline
    $b_i$           & $0.156$            & $0.116$             & $0.131$ & $0.201$ & $0.093$ & $0.303$ \\
    \hline\hline
  \end{tabular*}
  \caption{Characteristic values assumed for $b_i$ and $\tau_i$ in Eq.~(\ref{eqn:ianneal}), 
    based on details found in Refs.~\cite{Moll:1999kv,Wunstorf:1992ua}.  The time constants have been 
    scaled to various temperatures using the Arrhenius equation.}
  \label{tab:ianneal}
\end{table}

During annealing, the leakage current is observed to decrease and the
rate of this decrease strongly depends on temperature, based on
studies performed in the temperature range 0-60\degC.  The decrease of
the leakage current is often parameterized according to the Hamburg
model, which suggests a leakage-current evolution according to the
formula:
\begin{linenomath}
\begin{equation}\label{eqn:ianneal}
\Delta I(t) = \Delta I(t_0) \sum_i b_i \exp\left(-\frac{t}{\tau_i}\right) \ .
\end{equation}
\end{linenomath}
In this expression, $t_0$ represents the start time of annealing and
the sum is over different types of crystal defects, each of which has
a temperature-dependent characteristic time constant $\tau_i$ and an
amplitude $b_i$, subject to the constraint $\sum_i b_i = 1$.
Table~\ref{tab:ianneal} shows characteristic values for the constants
$b_i$ and $\tau_i$ for the annealing temperature of 18\degC, and also
11\degC and -5\degC, which are the the nominal operating temperatures
of the SVX and L00 CDF silicon subdetectors, respectively (see
Sec.~\ref{sec:det}).  Note that the weights $b_i$ are not temperature
dependent, but the time constants $\tau_i$ scale according to the
Arrhenius equation.  

\begin{figure}[bt]
\includegraphics[width=\textwidth]{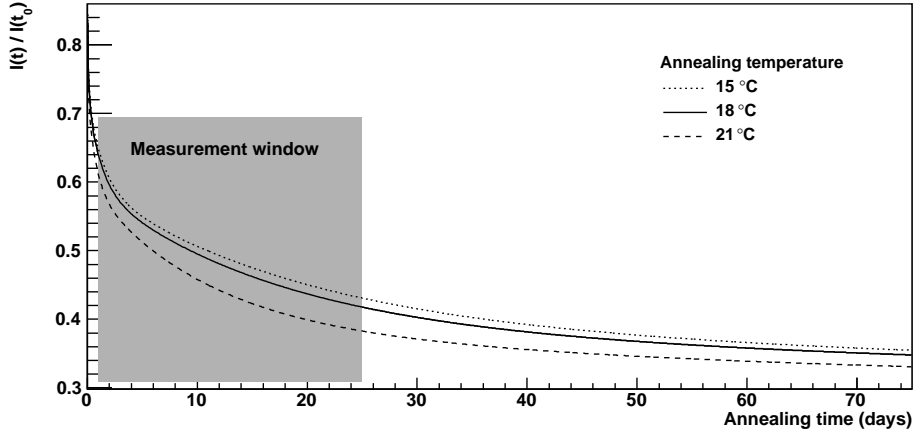} \\
\caption{Expected behavior of the bias current during annealing for
  various temperatures, based on constants in
  Ref.~\cite{Moll:1999kv,Wunstorf:1992ua}.  The shaded region
  corresponds to the period when annealing data was recorded for this study.}
\label{fig:annealCurrent}
\end{figure}

Figure~\ref{fig:annealCurrent} shows the expected leakage current
behavior during annealing for annealing temperatures of 15, 18 and
21~\degC.  As can be seen, the leakage current is at its maximum
immediately after warming, and then decreases due to the annealing
behavior as described in Eq.~(\ref{eqn:ianneal}).
The shaded region in Fig.~\ref{fig:annealCurrent} corresponds to the
period when CDF annealing data were recorded.  As the time constants
of the individual terms in Eq.~(\ref{eqn:ianneal}) span several orders
of magnitude, the measurements presented are sensitive to only a
subset of the parameters in Eq.~(\ref{eqn:ianneal}).  A more
appropriate parameterization is thus
\begin{linenomath}
\begin{equation}\label{eqn:iannealSimp}
\Delta I(t) = A_I \exp\left(-\frac{t}{\tau_I}\right) + B_I \ ,
\end{equation}
\end{linenomath}
where $A_I$ and $B_I$ are empirical constants, and $\tau_I$ is a time
constant associated with annealing; it is calculated to be 17.6 days,
as discussed in Sec.~\ref{sec:ibiasDiscussion}.

The change in the depletion voltage \vdep during annealing occurs in a
more complicated fashion.  According to the Hamburg model, as the
sensor is irradiated with an accumulated fluence $\Phi_{eq}$, \vdep
changes proportionally to any adjustments in the effective doping
concentration:
\begin{linenomath}
\begin{equation}
\Delta N_{\mathrm{eff}} = N_A(\Phi_{eq},t) + N_C(\Phi_{eq})+N_Y(\Phi_{eq},t)
\end{equation}
\end{linenomath}
where $t$ is the annealing time, and $N_A$, $N_C$ and $N_Y$ represent
contributions from short-term annealing, a stable damage component
independent of annealing time, and reverse-annealing, respectively. As
we are primarily interested in the time-dependence of annealing, $N_c$
merely serves as an overall offset, and so we do not specify its
explicit form.  The short-term and reverse annealing components,
are given by:
\begin{linenomath}
\begin{align}
N_A(t) & =  \quad \ N_A\exp \left(-\frac{t}{\tau_A}\right) \quad , \quad \quad \mathrm{and} \label{eq:na} \\
N_Y(t) & = \left\{\begin{array}{ll} 
    N_Y\left(1 - \exp\left(-k_{1Y}t\right)\right) & \mbox{for\ first-order\ process}\\ 
    N_Y\left(1-\displaystyle\frac{1}{1+k_{2Y}N_Yt}\right) & \mbox{for\ second-order\ process} \end{array}\right. \ .
\end{align}
\end{linenomath}
where the dependencies on the fluence $\Phi_{eq}$ have been absorbed
by the constants $N_A$ and $N_Y$.  An explanation of the definitions
and differences of first- and second-order processes can be found in
Ref.~\cite{Moll:1999kv}.

At room temperature, reverse-annealing has a time scale on the order
of 500 days~\cite{Moll:1999kv}, for which both first- and second-order
processes can be approximated for this analysis by a term linear in
annealing time: $N_Y(t)\approx N_Yt/\tau_Y$, where $\tau_Y$ is the
500-day time constant.  We therefore expect the depletion voltage
\vdep to follow
\begin{linenomath}
\begin{equation}\label{eqn:deltaV}
  \vdep = V_A\exp\left(-\frac{t}{\tau_V}\right) + V_C + V_Y\left(\frac{t}{\tau_Y}\right)
\end{equation}
\end{linenomath}
where $V_A$ and $V_C$ represent offsets, and $V_Y$ is the constant
associated with reverse-annealing.  The short-term annealing time
constant $\tau_V$ is expected to be $3.6^{+2.2}_{-1.3}$ days, based on
parameters given in Ref.~\cite{Moll:1999kv}, and scaling to 18~\degC
using the Arrhenius equation.  Note that the value of this time
constant is expected to be much less than that of
Eq.~(\ref{eqn:iannealSimp}).

\begin{figure}[!tb]
  \includegraphics[width=\textwidth]{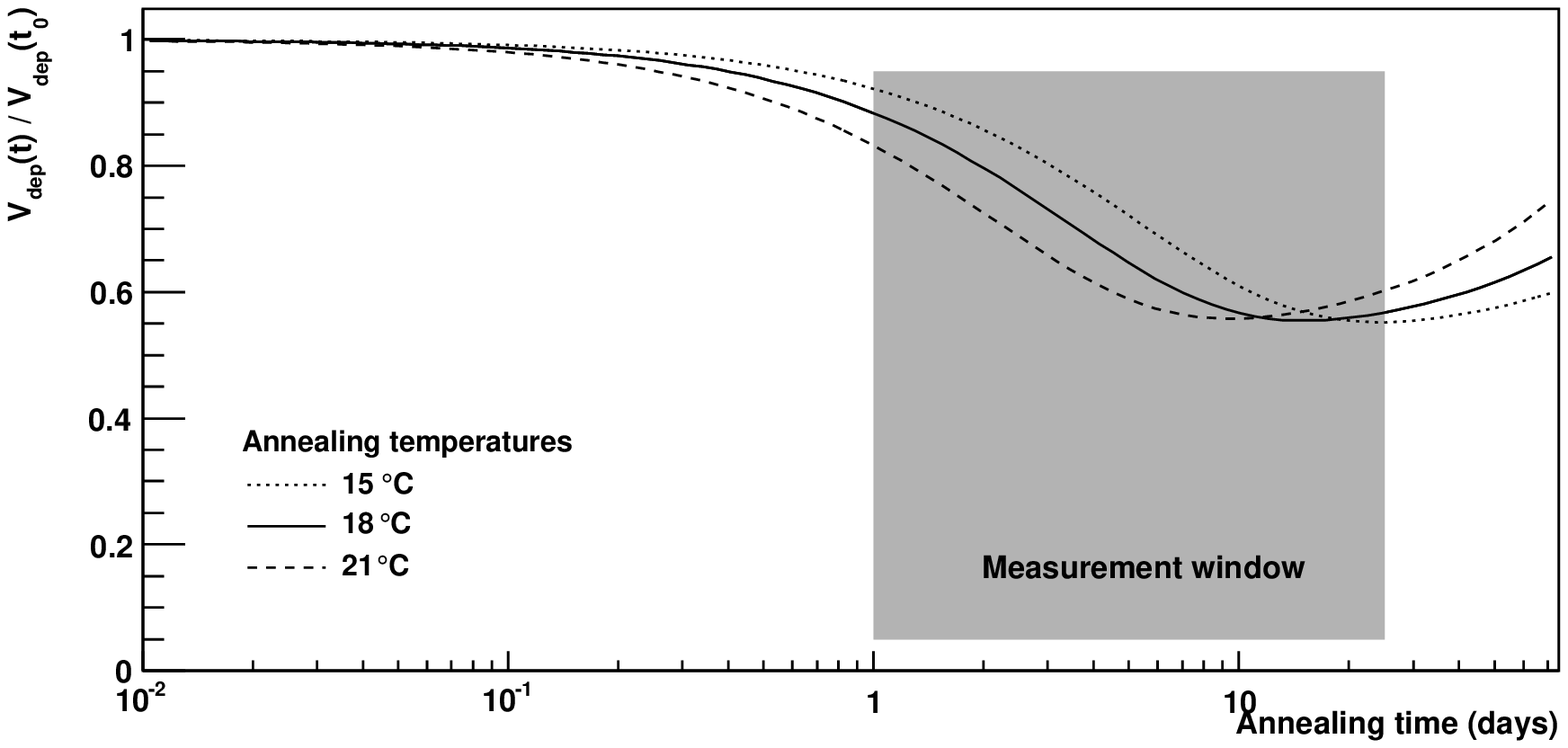} \\
  \includegraphics[width=\textwidth]{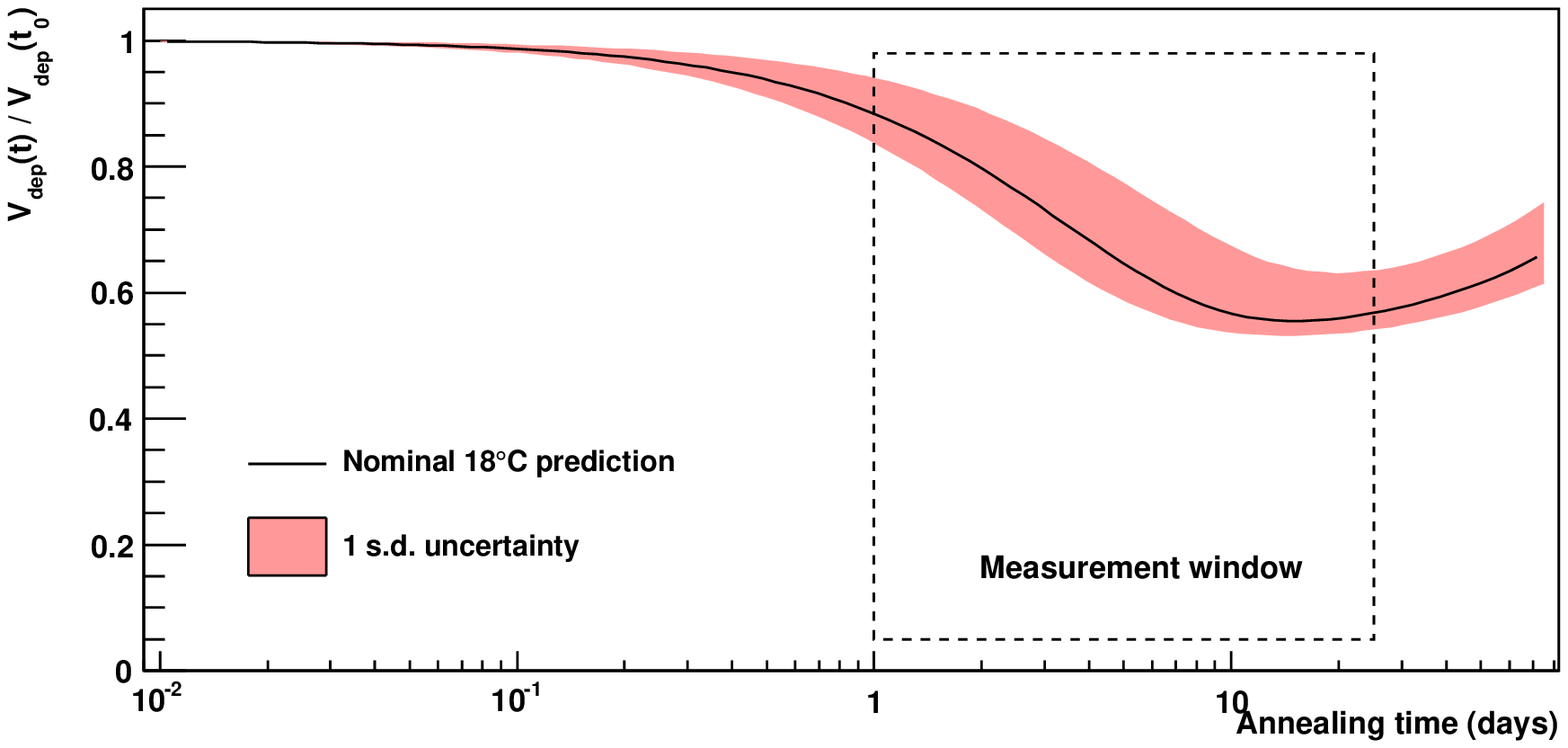}
  \caption{Expected behavior of the depletion voltage \vdep, assuming
    the Hamburg model, for various annealing temperatures (top), and
    for 18\degC, but including the one standard-deviation
    uncertainties on the parameters that are assumed in the model
    (bottom).  For this study, data were collected in the time period
    indicated by the boxed region.}
  \label{fig:vdepPredict}
\end{figure}

To illustrate the temperature dependence on the predicted annealing
behavior of \vdep, we plot the Hamburg-model prediction assuming
annealing temperatures of 15, 18, and 21~\degC, shown in the top plot
of Fig.~\ref{fig:vdepPredict}. For these predictions, we use values of
$V_A$, $V_C$ and $V_Y$ based on estimates made specifically for the
L00 narrow ladders of the CDF silicon detector (see
Sec.~\ref{sec:det}).  The lower plot of Fig.~\ref{fig:vdepPredict}
shows the nominal prediction for the \vdep annealing behavior at
18\degC, and also shows the uncertainty in that prediction, based on
estimates in the model parameters as derived in
Ref.~\cite{Moll:1999kv}.  Note that whereas \vdep is expected to reach
a minimum at some point during annealing, the same behavior is not
expected for the leakage current, which decreases monotonically as a
function of annealing time.

Assuming the Hamburg model appropriately describes the behavior of the
silicon sensors at CDF, we expect the after-before depletion voltage
ratio to reach a minimum before the end of the measurement window,
reaching a value somewhere between 50\% and 65\% assuming no annealing
prior to the end of Run II.  Due to scheduling constraints, we were
unfortunately unable to extend the measurement window to distinguish
between an asymptote and a minimum.


%% file: detector.tex
\section{The CDF Silicon Detector}
\label{sec:det}

\begin{figure}[h]
\centering
\includegraphics[width=0.47\textwidth]{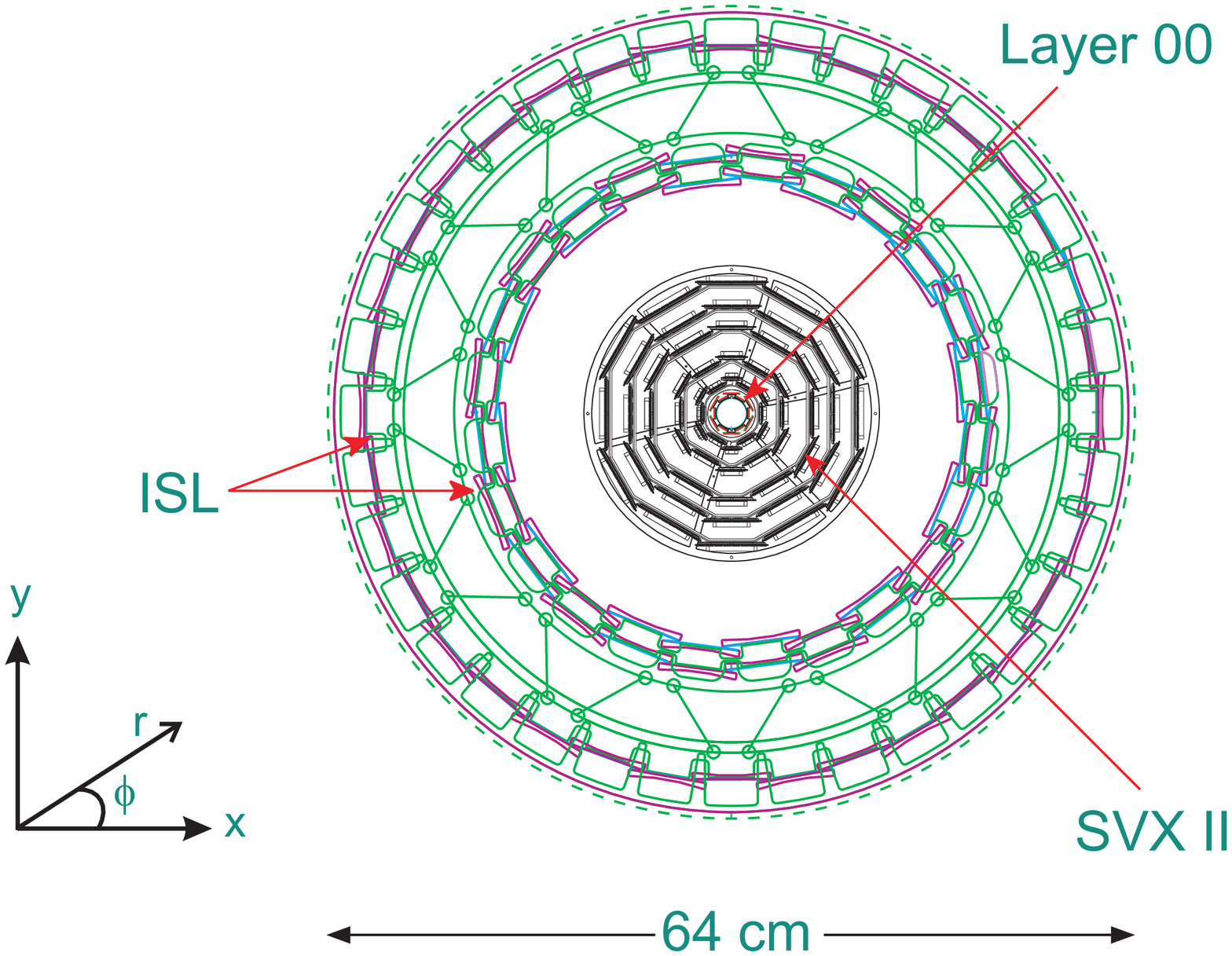}
\hspace*{0.04\textwidth}
\includegraphics[width=0.47\textwidth]{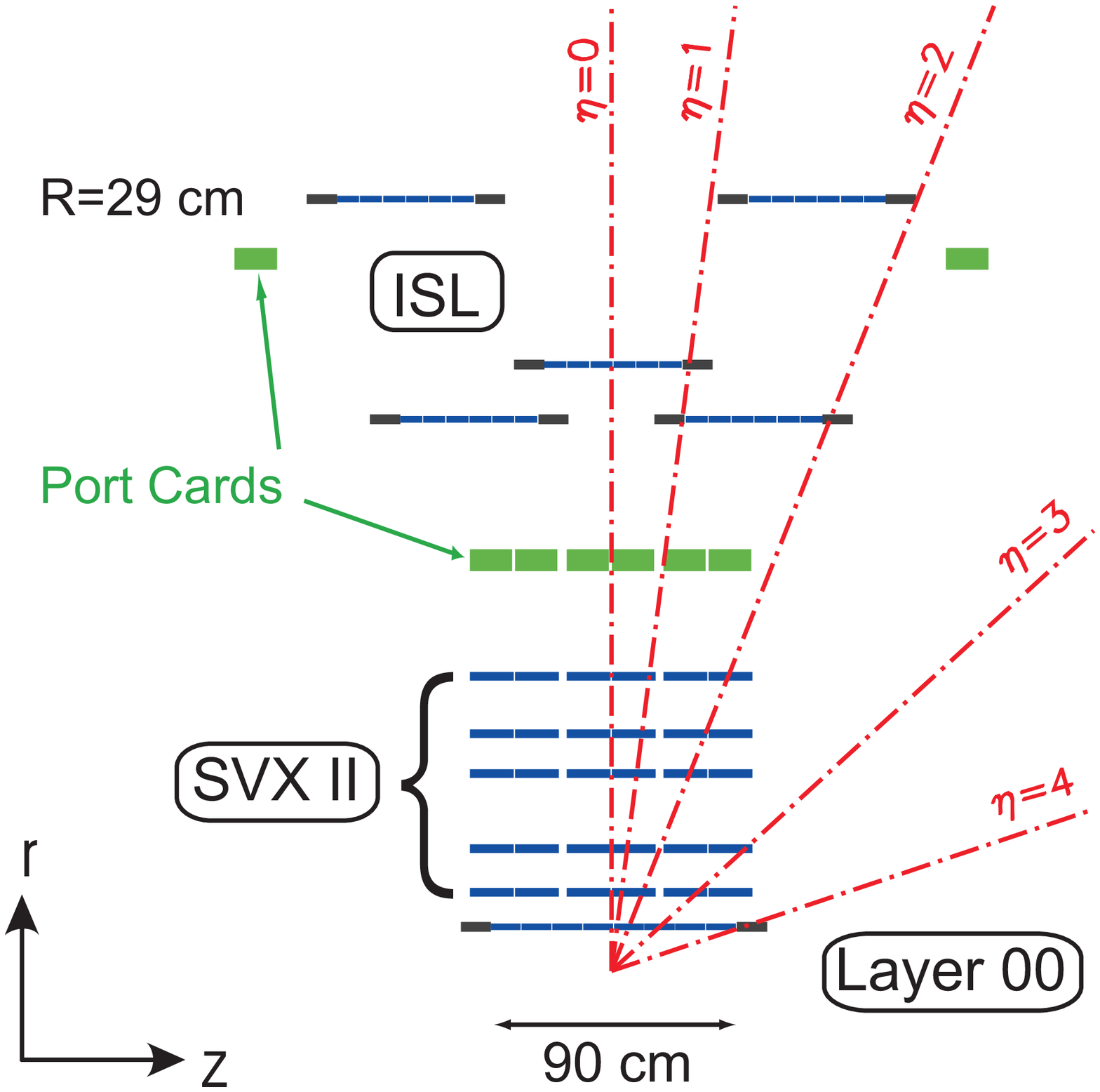}
\caption{Schematic layout of the CDF silicon detectors showing \rphi~
  (left) and \rz~(right) views. Note that the $z$ axis is compressed
  for illustration purposes.}
\label{fig:Sili_layout}
\end{figure}

The CDF silicon detector system~\cite{opsNIM} 
consisted of three sub-detectors, all
with barrel geometry: Layer 00 (L00), the
Silicon Vertex detector (SVX) and the Intermediate
Silicon Layers (ISL). Unless otherwise stated,
``detector'' refers to the CDF silicon detector.
Figure~\ref{fig:Sili_layout} presents the schematic layout of the CDF
silicon detector.
The design had eight silicon \textit{layers} to provide tracking which
was robust against failure or degradation of individual components.

%
%

The basic structural unit of a sub-detector was a \emph{ladder}, which
consisted of several silicon microstrip sensors bonded in series (3
sensors for L00 ladders, 4 in SVX ladders and 6 in ISL ladders).  Only
L00 sensors and SVX sensors in layers 0, 1 and 3 were used in this
study.  SVX sensors in layers 2 and 4 
developed complicated noise profiles making simple data analysis
described in this paper unfeasible. The ISL ladders were located too
far from the beamline to receive significant dose of radiation and
were therefore not of interest for this study, and are not
discussed further in this paper.

The sensors were made from high-resistivity $n$-type silicon with a
nominal thickness of 300 $\mu \mathrm{m}$. Sensors in L00 were
single-sided, providing \rphi\ information, while sensors in the other
layers were double-sided, providing both \rphi\ and \rz\ information.

A full ladder was read out from both ends through SVX3D readout chips
mounted on electrical hybrids, located outside (for L00) or inside
(for SVX) of the tracking volume.  Multiple readout chips were chained
together to read out a single silicon sensor.  A circuit board called
the \emph{portcard} was located at the periphery of each support
structure or bulkhead and formed an interface with the hybrids and
readout chips with the rest of the data acquisition system.

Layer 00 was a single-sided, single-layer silicon microstrip detector
whose sensors were grouped into 48 ladders.  It was mounted on a
carbon fiber support structure which in turn was mounted directly on
the beam pipe.  The L00 ladders are classified as \textit{narrow} or
\textit{wide}, based on the azimuthal extent of the ladder.  Narrow
ladders were positioned closest to the beam pipe at a radius of 1.35
cm; the wide ladders were located farther away from the beam pipe at a
radius of 1.62 cm.

The SVX detector was built in three cylindrical barrels each 29~cm
long. Each barrel contained five layers of double-sided silicon
microstrips placed along the beam axis, with radial coverage from 2.5
to 10.7~cm.  Carbon fiber reinforced Rohacell foam~\cite{Rohacell}
provided support to the ladders, and beryllium bulkheads provided
additional support and alignment on each end. Therefore the detector
consisted of six bulkheads ($z$-segmentation), each divided into 12
wedges ($\phi$-segmentation) consisting of 5 layers
($r$-segmentation).

The CDF silicon detector used power supply modules manufactured by
CAEN, model number A509 for SVX and model number 
A509H for L00.  These custom modules were
housed in SY527 mainframe crates located in the corners of the CDF
collision hall. One power supply module provided low voltages (2~V and
5~V) to the portcard, low voltages (5--8~V) to the SVX3D chip chains,
and high voltage (up to 500~V) to bias the sensors of one wedge of the
silicon detector. L00 ladders had two bias voltage lines. One voltage
line was connected to one of the three L00 ladder sensors while the
second one was used for the other two.

The SVX and L00 sensors shared a common cooling system to remove the
heat generated by the readout chips and maintain a constant operating
temperature for the silicon sensors.  The temperature of the coolant
exiting the chiller was $-10$\degreesC. The SVX sensors were not in
close thermal contact with the coolant and their temperatures were not
directly monitored.  However, by combining the measurements of the
ambient and coolant temperatures
as a function of time with a finite element thermal model and a
dedicated post-run measurement, the temperatures of the sensors during
data-taking conditions are estimated to be between 10 and 12\degC for
SVX.  For the L00 sensors, cooling was achieved through thermal
contact to aluminum tubes glued to the mechanical supports.  The L00
readout chips were distant from the sensors, and cooled by separate
cooling lines, allowing a temperature of -2.5\degC for the L00 sensors
during data taking operations.

The radiation dose the detector was exposed to was estimated using TLD
measurements of the radiation field in the CDF tracking
volume~\cite{TLD}, extrapolated to the location of the individual
silicon layers. The equivalent dose from 1~MeV neutrons can be
approximated by assuming that the contributions from photons and
low-energy neutrons to the TLD measurements are negligible and that
the damage is caused primarily by high-energy charged pions.  This
results in the relations 1~Mrad $\approx 3.9\times10^{13}$~high-energy
pions/$\mathrm{cm}^2~\approx2\times10^{13}$~1~MeV
neutron/$\mathrm{cm}^2$ equivalent.

Table~\ref{tab::sensorsummary} provides information about the number
of ladders in each layer used in this measurement, distance from the
beamline, estimated radiation dose, as well as operating voltage and
bias current at the end of Run II. The bias currents in sensors from
the same layer vary by $20$\%, largely due to temperature differences
among the sensors. It is worth noting that only one sensor per L00
ladder was used in this measurement because the bias voltage lines
connected to two sensors drew too much current to be powered at
18\degC.

\begin{table}[bt]{
\footnotesize
\begin{tabular*}{\textwidth}{@{\extracolsep{\fill}}l|ccccc}
  \hline \hline
                                  & L00N        & L00W        & SVX-L0          & SVX-L1          & SVX-L3           \\ \hline
  Number of ladders               & 12          & 36          & 72              & 72              & 72               \\
  Sensors per ladder              & 3           & 3           & 2               & 2               & 2                \\
  Distance from detector axis     & 1.35~cm     & 1.62~cm     & 2.54~cm         & 4.12~cm         & 8.22~cm          \\
  Expected dose                   & 11.5~Mrad   & 8.7~Mrad    & 4.5~Mrad        & 2.2~Mrad        & 0.76~Mrad        \\
  Average \ibias per sensor       & 200~$\mu$A  & 250~$\mu$A  & 500~$\mu$A      & 400~$\mu$A      & 300~$\mu$A       \\
  Power supply limit per line     & 3000~$\mu$A & 6000~$\mu$A & 5000~$\mu$A     & 5000~$\mu$A     & 5000~$\mu$A      \\ 
  Sensor temperature              & -3~\degC    & -3~\degC    & 11~\degC        & 11~\degC        & 11~\degC         \\ 
  Max. operating voltage          & 500 V       & 500 V       & 200 V           & 200 V           & 200 V            \\ 
  Final operating voltage         & 365 V       & 300 V       & 170 V           & 100-130 V       & $<$100 V         \\ 
  \hline \hline
\end{tabular*}
\caption{Number of ladders in each layer, distance from the beamline, estimated 
  radiation dose, as well as operating voltage and bias current 
  at the end of Run II. The L00 ladders are classified as L00N or L00W, based on 
  the whether the ladder is narrow or wide in azimuthal extent, respectively. The current temperature
  of L00 has a 2.5~\degC uncertainty and/or variation.  The current temperature of SVX has a 5~\degC
  uncertainty and/or variation among the sensors.}
\label{tab::sensorsummary}
}
\end{table}

%% file: procedure.tex
\section{Measurement Procedure}
\label{sec:meas}
The annealing measurement presented in this paper lasted for 24 days.
The results are based on the measurements of $IV$ curves of the
sensors and the evolution of these curves as a function of annealing
time. For this study, the detector was warmed up from operating
temperature described in the previous section to 18\degC. Higher
temperatures were desired for the annealing measurement, but could not
be reached due to current in the bias lines exceeding the power supply
limit.  Because the bias current is very sensitive to the temperature
of the ladder, attempts have been made to minimize the effects of
ladder self-heating and cross-heating from other ladders used in the
study.  Unless otherwise stated, detector/ladder OFF state in this
paper refers to both high and low voltages set to 0 volts and
detector/ladder ON state refers to both high and low voltages set to
their nominal values.

\subsection{Power Supply Modifications}
As stated earlier, CAEN power supply modules provided low voltages to
the portcard, low voltages to the SVX3D chip chains, and high voltage
to bias the sensors of one wedge of the silicon detector.  A safety
feature prevented the modules from powering ON the high voltage (HV)
channels without first ensuring that the corresponding low voltage
(LV) channels are ON. Another safety feature of the power supply
modules prevented them to be powered ON when one or more HV or LV
cables are disconnected.  In other words, in the Run II operating
configuration, it was impossible to apply bias voltage without
switching on power to the portcards and readout chip chains.  When
powered, these electronics provided significant heat to the sensors.
Therefore, modifications were needed to the power module configuration
in order to decouple the HV and LV channels and apply bias voltage
with LV cables disconnected.


\subsection{$IV$ scan software}
Custom software was developed in order to perform automatic $IV$
scans. A scan consisted of varying the bias voltage from 0 to
$V_{max}$ in multiple steps for a particular ladder. The value of
$V_{max}$ depended on the detector layer to which the ladder belonged.
The step size was also layer dependent and typically in the range of
5--10 V. Not more than two L00 ladders were scanned simultaneously to
avoid the effects of cross-heating. Moreover, any ladders scanned in
parallel were required to be well separated in the detector volume.
For such configurations, the effects of cross-heating were proved to
be negligible by observing no change in the bias current for one of
the ladders while the other was powered on and off.

\subsection{Detector Monitoring}
\label{sec:mon}
Monitoring software was developed to 
ensure successful execution of the annealing study. The \iv, 
\ibias-\textit{vs.}-time and \vbias-\textit{vs.}-time curves were
stored for each scan and checked each day.  Any changes in operating
temperature outside the allowed tolerance triggered an alarm.

\subsection{Timeline of the Measurement}
\label{sec:timeline}
Preparation for the measurements started on 09/30/2011, the official
end of the Tevatron Run II.  
Dedicated $IV$ scans with L00 and SVX low voltage ON were performed.
These data help determine the overall change in depletion voltage
during the annealing period.  L00 and SVX were switched OFF after
these scans, and the chiller set point temperature raised to -5\degC to
avoid freezing in the ISL cooling pipes.

On 10/3/2011, the modifications to the power supply modules were
completed, and the LV cables were left disconnected from the power
supplies until the final day of the study.
In parallel, the interlock system settings were changed to allow
powering up the detector at temperatures higher than allowed during
data taking.  The warm-up started on 10/4/2011 and was performed by
raising the chiller set point temperature in three steps: to 5\degC,
to 15\degC, and to 18\degC. The warm-up stages were separated by
three hours in time to allow the temperature in the detector volume to
stabilize. $IV$ scans were performed at the end of each stage of the
warm-up.

From 10/4/2011 to 10/27/2011, stable thermal conditions were
maintained, except for two malfunctions of the chiller that regulated the
coolant temperature.  These resulted in colder temperatures for
2-3 hours, and data acquired during these periods were discarded.
$IV$ scans were performed on groups of 2 L00 ladders at a time, with
each ladder being scanned roughly every 21 hours.  The temperature of
the detector volume and the status of the power supply modules were
closely monitored.

On 10/28/2011, the chiller set point was lowered from from 18\degC to
9\degC to measure bias currents of the SVX ladders at a controlled
temperature, uniform across the ladders.  These data were used to
determine the operating temperature of the SVX ladders reported in
Table~\ref{tab::sensorsummary}.  On 10/30/2011, the chiller set point
was lowered to 0\degC to provide stable and uniform thermal conditions
for the L00 ladders.  The bias current of each L00 ladder was measured
both at the full operating voltage and half the operating voltage.
Only one ladder was powered at a time for maximum thermal stability.
These data were used to determine the thermal coupling constants
$\kappa$ for each ladder, needed for the self-heating corrections
discussed in \ref{sec:sh}.

Finally, on 10/31/2011, the LV cables were reconnected to the power
supply modules, the chiller set point lowered to -10\degC, and the LV
power turned on. Another set of $IV$ scans under operating (data
taking) conditions was recorded.  The change in depletion voltage over
the annealing period was determined by comparing these scans with
those taken before the warm up with the same thermal conditions.

%% file: currents.tex
\section{Bias Current Evolution}
\label{sec:curr}
 
\begin{figure}
\centerline{\includegraphics[scale=0.5]{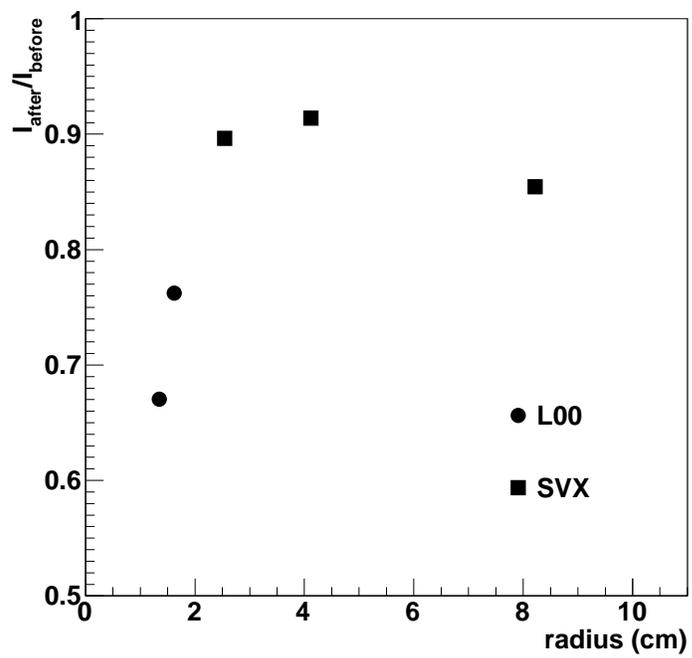}}
\caption{The ratio of the bias current after the
annealing to that before the annealing, averaged over all functional ladders 
in a single layer, is plotted as a function
of the distance of the ladders from the central axis of the detector.}
\label{fig:currratio}
\end{figure}

The change in bias current during the annealing period in two ways.
We first measure the fractional change over the entire annealing
period by comparing the bias current measured under operational
thermal conditions before and after the annealing period.  Under these
thermal conditions, the self heating of the ladders is negligible.
Figure~\ref{fig:currratio} shows
$I_{\mathrm{after}}/I_{\mathrm{before}}$, the ratio of the bias
current after the annealing to that before the annealing, as a
function of radial distance from the beam axis.  The average is taken
over all functional ladders in the layer.

Secondly, we track the bias current evolution over the course of
annealing.  By examining the current measured at the largest voltage
of each IV scan, roughly every 21 hours, the shape of the bias current
decrease can be examined.  The self-heating of the sensors is
substantial under these circumstances, increasing the temperature as
much as 3\degC.  The bias currents have a strong temperature
dependence, increasing roughly 10\% for every degree increase in
temperature~\cite{sze1985}.  Thus, the measured currents must be
corrected back to the equivalent current at 18.2\degC before
information about the annealing processes can be extracted.  It is
assumed that the temperature increase is linearly proportional to the
power dissipated by the ladder, and the constant of proportionality
$\kappa$, unique to each ladder, was determined with a dedicated
measurement described in \ref{sec:sh}.  Only the narrow
ladders of L00 are used for the warm-temperature measurements.  This
is because the wide ladders dissipate 3-4 times more heat due to their larger
sensor volume, and they have a weaker coupling to the cooling system,
which introduces complications to the self-heating correction
procedure.

\begin{figure}
\centerline{\includegraphics[scale=0.5]{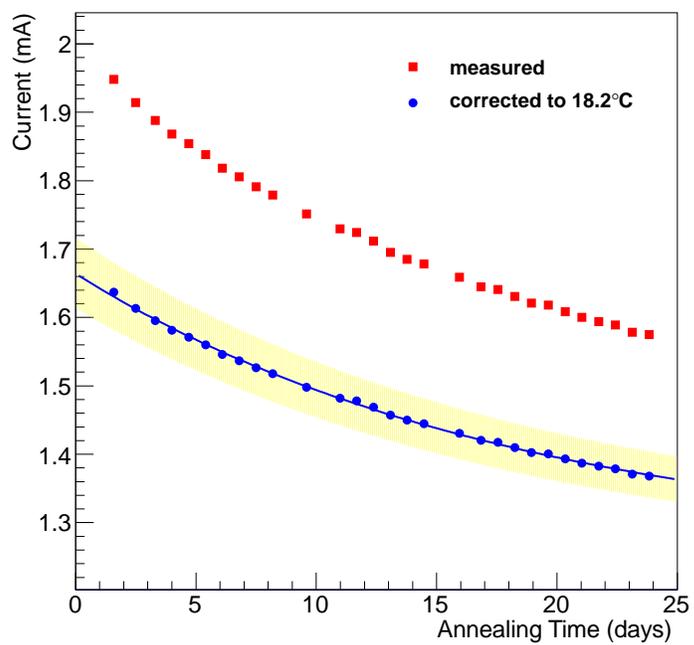}}
\caption{The time constant of this fit is $18.29 \pm
  1.24_{-0.20}^{+0.22}$ days.  The first uncertainty is statistical
  and the second systematic, derived from from the uncertainty on
  $\kappa$.}
\label{fig:narrowCurr}
\end{figure}

Figure~\ref{fig:narrowCurr} shows the data for a typical L00 narrow
ladder.  The measured currents at the full voltage are shown as a
function of annealing time with red squares.  The blue circles
represent the equivalent current at 18.2\degC.  The annealing time
dependence of the corrected currents is fit to
Eq.~(\ref{eqn:iannealSimp}), and the fit result is shown as a solid
line.

An alternate fit function, with an additional exponential term, was
considered.  However, the uncertainties on the parameters of the
additional term were large, suggesting that it is not needed to
describe the data.  This is discussed further in
Sec.~\ref{sec:results}.

The uncertainty on the parameter $\kappa$ in the temperature
correction determines the shaded uncertainty band shown around the
best fit in Fig.~\ref{fig:narrowCurr}, and is used to determine the
systematic uncertainty on the time constant of the decaying
exponential.  A compilation of the fit results for all ladders is
presented in Sec.~\ref{sec:results}.

%% file: dv.tex
\section{Depletion Voltage Changes}
\label{sec:dv}

As mentioned in Sec.~\ref{sec:si_an}, the behavior of a silicon sensor
is partially characterized by the depletion voltage \vdep, which is
the minimum bias voltage that depletes the active detector region of
any charge carriers.  In the context of test beam setups, the value of
\vdep is determined by measuring the capacitance of the sensor as a
function of \vbias.  The $CV$ curve exhibits a kink at high voltage,
and the intersection of two lines that describe the data before and
after the kink unambiguously defines \vdep.  For an operating
experiment, the capacitance cannot be measured directly (as in a test
setup for a bare sensor), so an
alternate definition of \vdep is used, which corresponds to the
voltage that maximizes the signal collection in a given
data-collection time window (see Ref.~\cite{opsNIM} for details).

For the annealing study presented here, a signal source (i.e. source
of charged particles) was not available, thus requiring an alternate
method to determine \vdep.  As the shape of the \iv curve is the only
feature from which we can infer the internal properties of the silicon
sensor, we develop a metric called \vknee, which is a quantity extracted
from a fit to the \iv characteristic itself, as discussed in
Sec.~\ref{sec:ivcurvedata}.

We use \vknee to measure two quantities:
\begin{enumerate}
\item The relative change in depletion voltage, based on the mapping
  method described in Sec.~\ref{sec:dvmap}, and
\item The evolution of \vknee throughout the annealing process,
  characterized by a time constant as described in Sec.~\ref{sec:kv}.
\end{enumerate}

\subsection{\iv Fit Procedure}
\label{sec:ivcurvedata}

For each \iv scan taken, the data are fit to a function, motivated by
the Shockley formula for a $p$-$n$ junction~\cite{sze1985}, and an
additional term linear in \vbias, which accounts for radiation-damage
effects, approximated by a resistor placed in parallel with a $p$-$n$
junction:
\begin{linenomath}
\begin{equation}\label{eqn:fit}
\ibias(\vbias;\mathbf{p}) = p_0 - p_1\exp\left(-p_2\vbias^{p_3}\right) + p_4\vbias \ ,
\end{equation}
\end{linenomath}
where the $\{p_i\}$ represent parameters to be fitted.  The \iv data
are fit using 80\% of the voltage range, so as to minimize potential
residual self-heating effects that can occur at the largest voltages.
The \ibias uncertainties assumed correspond to half of the spread of
the measured bias current for a given voltage setting, after effects
from self-heating have stabilized.  Typically this uncertainty is on
the order of a few \uA.  An additional uncertainty of 1 \uA,
corresponding to the uncertainty of the least-significant bit, is
added in quadrature to this spread.

We define the knee voltage \vknee as the voltage where the slope of
the fit reaches 5\% of its maximum value, relative to the difference
of the maximum and minimum slopes.  The uncertainty in \vknee is
determined by propagating the uncertainties on the fit parameters
(assumed to be Gaussian-distributed about their central values), using
a pseudo experiment study that accounts for the fit-parameter
correlations.

\begin{figure}
  \includegraphics[width=\textwidth]{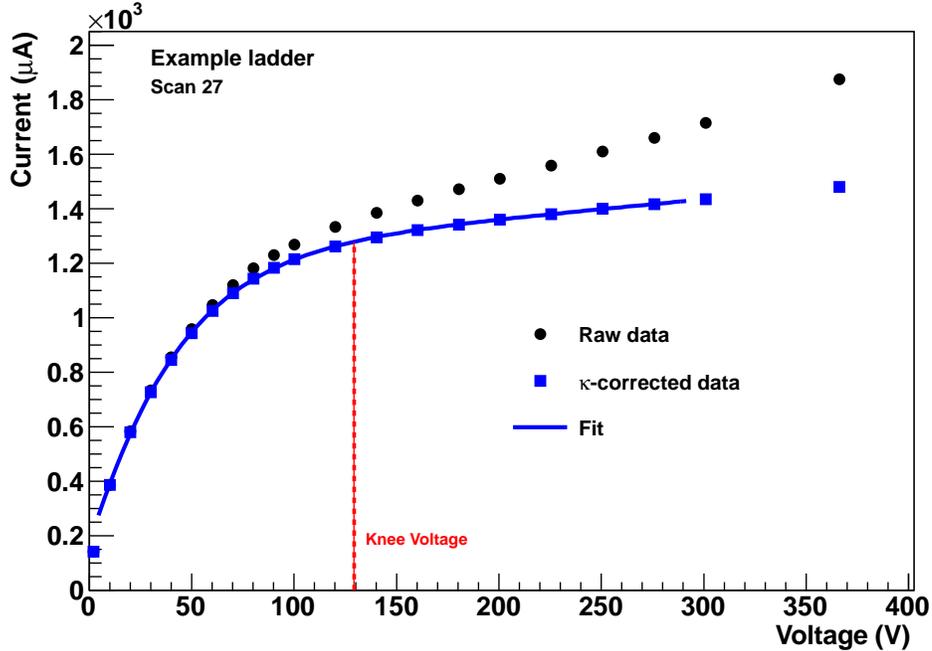}
  \caption{The fit of a warm \iv scan for an example ladder, corrected
    for self-heating effects.  Also shown are the uncorrected data
    points in black.}
  \label{fig:ex_scan27_cent}
\end{figure}

Figure~\ref{fig:ex_scan27_cent} shows an example fit for one of the
L00 ladders.  The original, uncorrected data points are shown as black
circles, and the $\kappa$-corrected points as blue squares.  The \iv
fit is performed on the $\kappa$-corrected \iv points, using the
parameterization of Eq.~(\ref{eqn:fit}).  Also shown is the extracted
knee voltage, and its associated uncertainty (which is on the order of
a few V, and thus difficult to see for this particular fit).  As
illustrated in Fig.~\ref{fig:ex_scan27_cent}, the effect of the
self-heating correction $\kappa$ (\ref{sec:sh}) can be significant at
higher voltages.  This behavior is not observed for low-voltage
ON \iv scans, which were performed at low temperatures.  The $\kappa$
corrections are thus omitted for the analysis of the LV ON scans,
which are described in Sec.~\ref{sec:dvmap}.

\subsection{Overall Relative Change in \vdep}
\label{sec:dvmap}

As mentioned in Sec.~\ref{sec:timeline}, separate low-voltage (LV) ON
scans, corresponding to operating conditions, were taken before and
after the annealing period. Although we have no signal source to infer
\vdep values for these scans, we develop a map between \vknee and
\vdep for the L00 ladders. To construct such a map, we use information
from the scans performed during nominal CDF running that determined
the actual depletion voltage \vdep, and the corresponding \vknee value
derived from fits to the \iv data using the same functional form as
shown in Eq.~(\ref{eqn:fit}).  We take the extracted \vknee values and
plot them against the measured \vdep values, and then use an
analytical expression to relate the two sets of values.

\begin{figure}[bt]
  \includegraphics[width=0.48\textwidth]{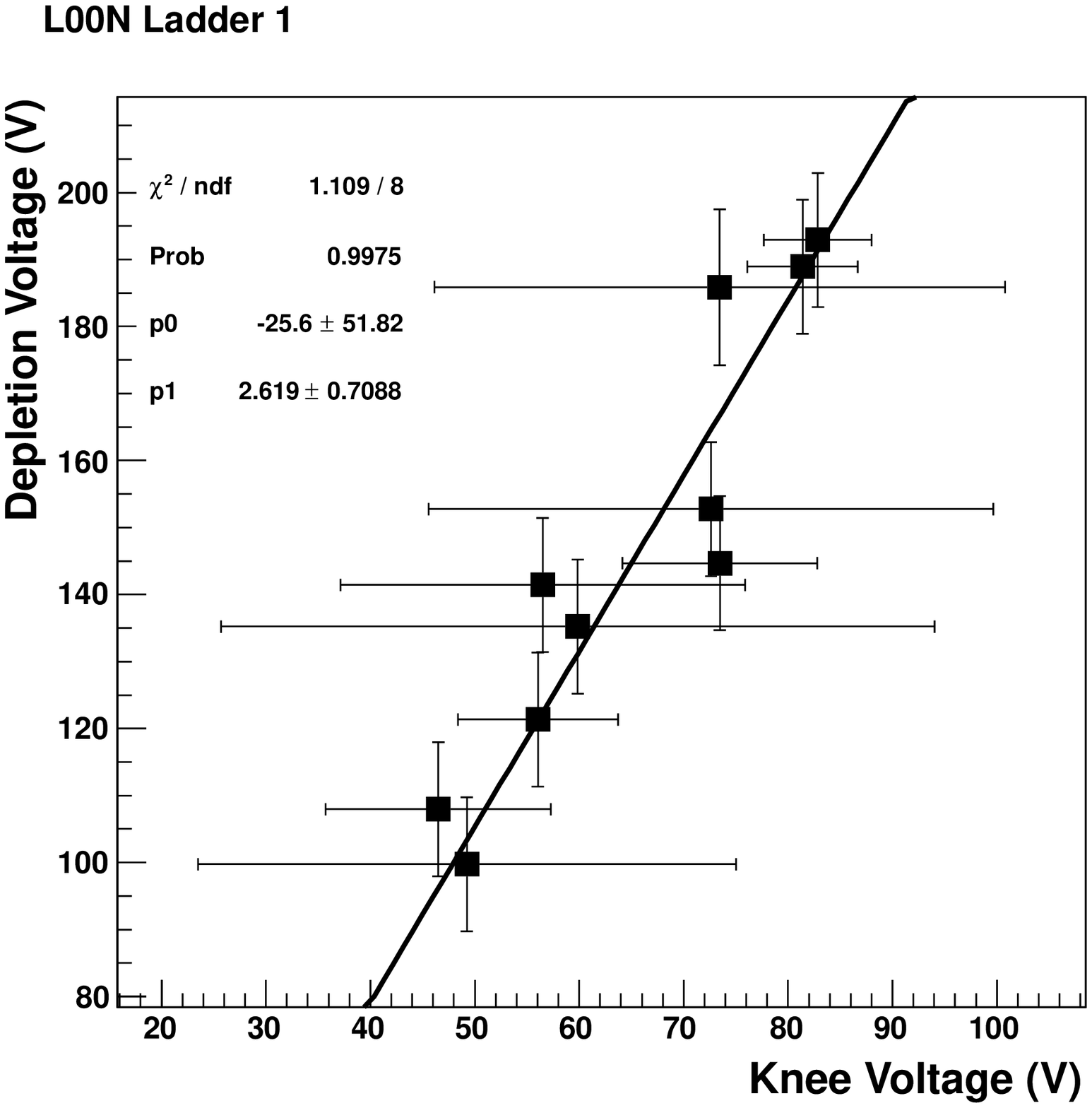}\hfill 
  \includegraphics[width=0.48\textwidth]{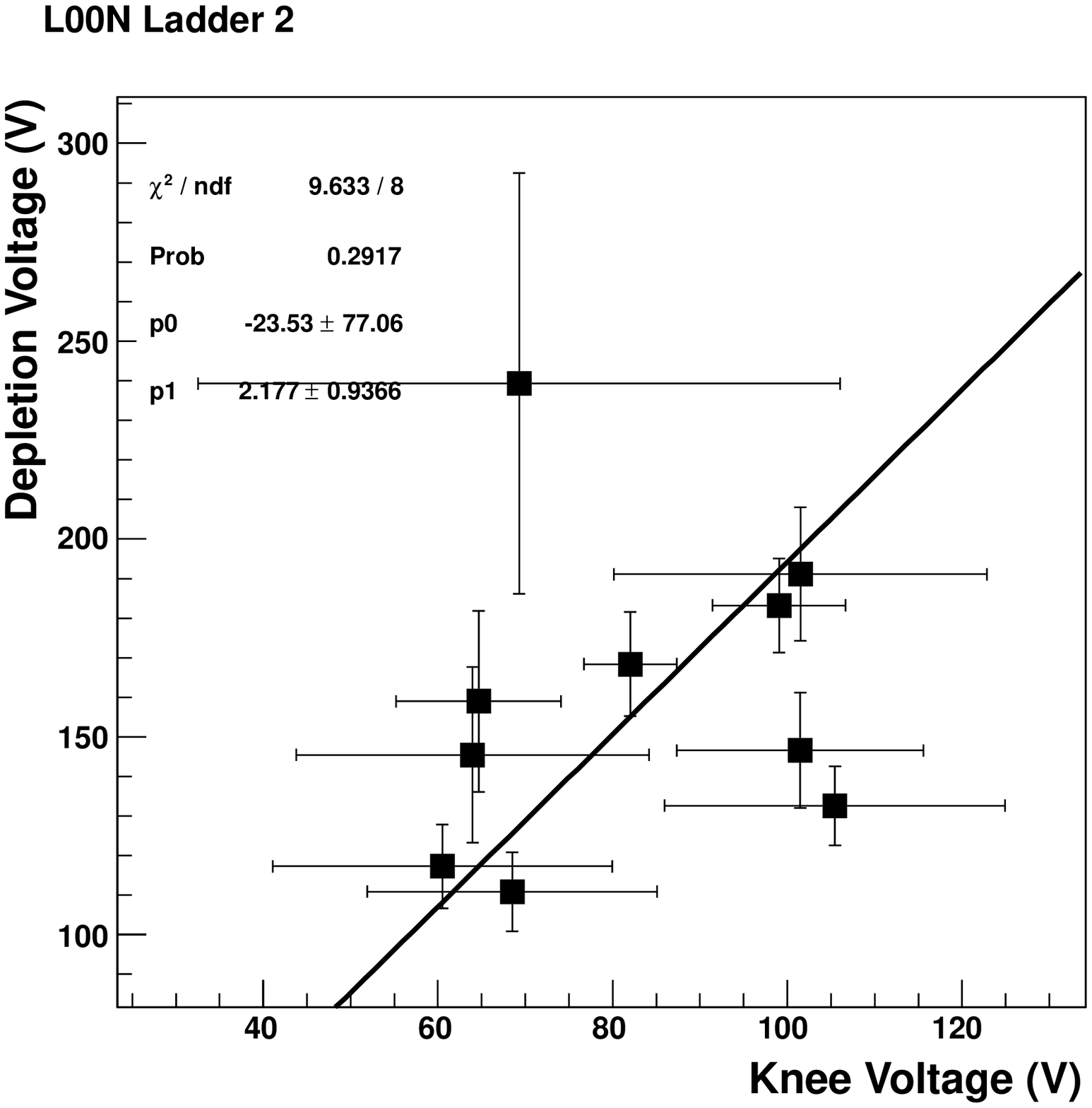}
  \caption{\vknee-\vdep mappings for two L00 narrow ladders.  The
    uncertainties in both directions are taken into account using the
    ``effective variance'' fit method.  In general, the $\chi^2$/ndf
    is very consistent with unity. The plot on the right shows the
    poorest example of any obvious mapping between \vdep and \vknee.}
  \label{fig:map}
\end{figure}

Whereas the \vknee determination of each nominal-running scan is
usually reliable, the \vdep value often includes large uncertainties.
This is a result of increased radiation damage on the silicon ladder
as a function of integrated luminosity.  We therefore assume a simple
linear parameterization $\vdep = p_0 + p_1\vknee$, which associates a
given value of \vknee to \vdep.  Some sample mappings are given in
Fig.~\ref{fig:map}.

Using these linear mappings, we take the measured \vknee values from
the LV ON scans before and after annealing and associate them with
mapped depletion voltages \vdepmap.  We then infer the decrease in
mapped depletion voltage, by forming the ratio $\vdepmap/\vdepmap^0$,
where $\vdepmap^0$ corresponds to the mapped depletion voltage before
annealing.  This quantity is measured for eleven L00 ladders, and the
result from each ladder is combined into a global average, presented
in Sec.~\ref{sec:results}.

Note that by using these mappings, we assume that:
\begin{enumerate}
\item \vknee serves as a reliable metric of the \iv curves that
  associates the \iv data to a unique value of the depletion voltage
  \vdep, and
\item as a mapping can be made only of pre-annealed data, we assume
  that the behavior of \vknee before annealing corresponds to its
  behavior afterward.
\end{enumerate}

\subsection{\vknee Evolution}
\label{sec:kv}

Although it would have been desirable to track the evolution of
\vdepmap over the course of the full month of annealing, the
\vknee-\vdep maps as described in Sec.~\ref{sec:dvmap} cannot be used
due to the non-trivial temperature dependence of \vknee.  To use the
maps, the silicon sensor temperatures would have had to be lowered to
nominal-running temperatures before each \iv scan was taken, which was
impractical.  Instead, we track the evolution of the knee voltage
during the annealing period.

\begin{figure}[bt]
  \centerline{\includegraphics[width=\textwidth]{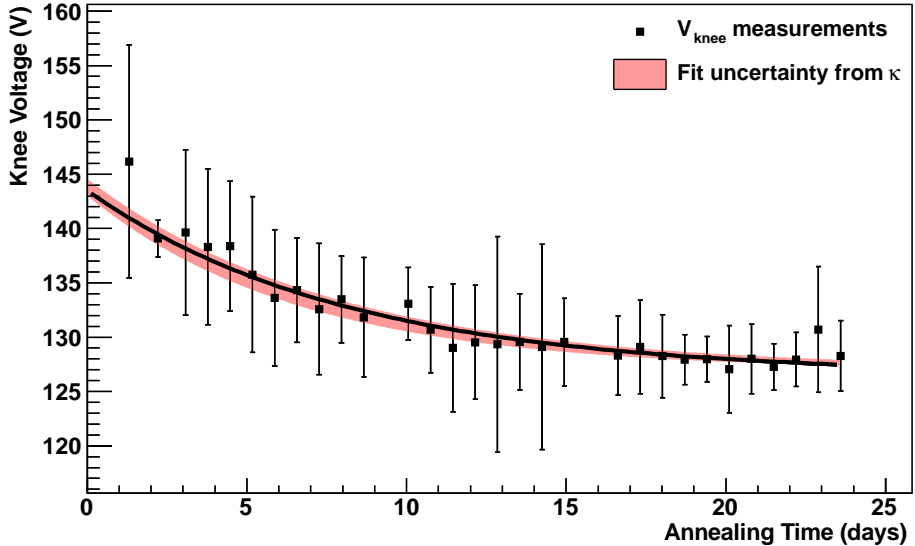}}
  \caption{Exponential fit to \vknee values for a typical L00N ladder.
    Variations in the fit due to the uncertainty on $\kappa$ are shown
    as a red, shaded band.}
  \label{fig:vkneefit} 
\end{figure}

Figure~\ref{fig:vkneefit} shows the \vknee values, associated with
fitting the \iv curves over the course of a month, for one of the L00
ladders. To extract an overall time constant, we assume the \vknee
evolution follows Eq.~(\ref{eqn:deltaV}), and we fit for $V_A$, $V_C$,
and the time constant $\tau_V$---we omit the reverse-annealing term
proportional to $V_C$ as we observe no increase in \vknee at larger
values of annealing time.  The uncertainties in the \vknee values used
in the fit are likely correlated between measurements as a function of
annealing time.  However, as this level of correlation is unknown, we
assume them to be uncorrelated, which translates to a larger
uncertainty in $\tau_V$. The solid line in Fig.~\ref{fig:vkneefit}
corresponds to the best fit using the Eq.~(\ref{eqn:deltaV})
parameterization without the reverse-annealing term, resulting in a
fitted time constant of $\tau_V = 6.14 \pm 0.38$ days.

Systematic uncertainties due to the self-heating correction $\kappa$
are incorporated by varying $\kappa$ by its uncertainty in the
positive and negative directions, and refitting the $\kappa$-corrected
\iv data.  The fits corresponding to varying $\kappa$ are shown as a
red, shaded region in Fig.~\ref{fig:vkneefit}.  From these alternate
fits, we extract the corresponding time constants and assign the
maximum deviation from $\tau_V$ as the systematic uncertainty on the
result.  For the L00 ladder shown, the systematic uncertainty is 0.07
days, giving $\tau_V = 6.14 \pm 0.38$(stat.)$\pm 0.07$(syst.)~days.
This analysis is repeated for all narrow ladders.  A weighted average
of the resulting time constants is performed and shown in
Sec.~\ref{sec:results}.

%% file: results.tex
\section{Results of the Annealing Study}
\label{sec:results}

\begin{figure}[bt]
  \centering
  \includegraphics[width=0.7\textwidth]{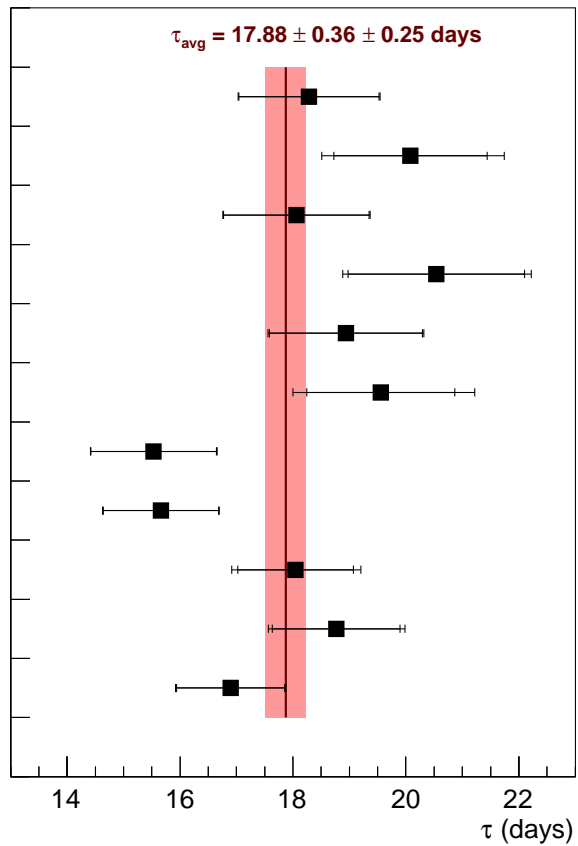}
  \caption{Time constants for decrease in temperature-corrected
    current corresponding to \vop.}
  \label{fig:timeconstants_current}
\end{figure}

Figure~\ref{fig:timeconstants_current} displays the fitted time
constant of the bias current evolution shape for each of the narrow ladders,
as described in Sec.~\ref{sec:curr}.  The individual results are
consistent with a single time constant, suggesting that potential
differences among the ladders due to (e.g.) annealing temperature or
radiation dose variations are small.  The weighted average of
$\tau_I=17.88\pm0.36\pm0.25$~days is obtained by assuming uncorrelated
statistical uncertainties and fully correlated systematic
uncertainties due to the self-heating corrections, as described in
\ref{sec:sh}.  The total uncertainty on the combined result is
indicated with a shaded band.

\begin{figure}[bt]
  \centering
  \includegraphics[width=0.7\textwidth]{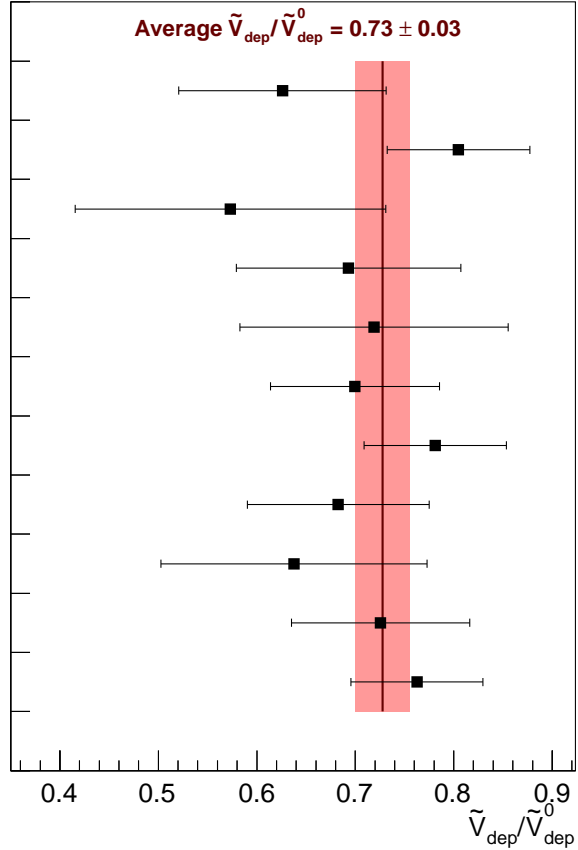}
  \caption{$\vdepmap/\vdepmap^0$ determinations for the eleven narrow
  L00 ladders, along with the weighted average.}
  \label{fig:avgDepVoltRed}
\end{figure}

Figure~\ref{fig:avgDepVoltRed} shows the after-before ratio in mapped
depletion voltage \vdepmap, as described in Sec.~\ref{sec:dvmap}, for
each of the L00 ladders considered as well as a weighted average. The
uncertainties of each \vdepmap determination are assumed to be
uncorrelated.  The average after-before \vdepmap ratio is thus $0.73
\pm 0.03$, indicating an average reduction of roughly 25\% in \vdep
due to annealing.

\begin{figure}[bt]
  \centerline{\includegraphics[width=0.7\textwidth,clip]{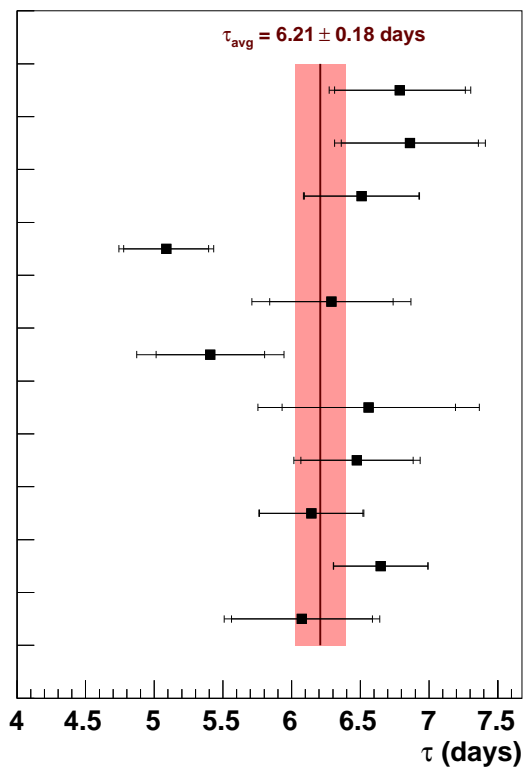}}
  \caption{Annealing time constants for the eleven narrow L00 ladders,
  and the weighted average, assuming the measurements are
  uncorrelated.}
  \label{fig:timeconstants_kv} 
\end{figure}

Finally, we present the averaged result for the \vknee evolution time
constant in Fig.~\ref{fig:timeconstants_kv}, as described in
Sec.~\ref{sec:kv}.  The results is an average of $\tau_{\mathrm{avg}}
= 6.21 \pm 0.18$ days. The statistical uncertainties are again assumed
to be uncorrelated, whereas the systematic uncertainties due to
$\kappa$ are treated as fully correlated in the weighted average.  The
solid vertical line and shaded region represent the weighted
average and its total uncertainty, respectively.

\section{Comparison with the Hamburg Model}
\label{sec:results2}

In order to compare these measurements with the Hamburg Model, the
fraction of annealing that happened during the run must first be
estimated using the record of sensor temperatures as a function of
time. The temperatures of SVX and L00 sensors during data taking
operations were roughly 11\degC and -2.5\degC, respectively.  During
shutdowns, with the heat load from the readout chips removed from the
system, temperature of both the SVX and L00 sensors was 0-5\degC.

\begin{figure}
\centerline{\includegraphics[width=\textwidth]{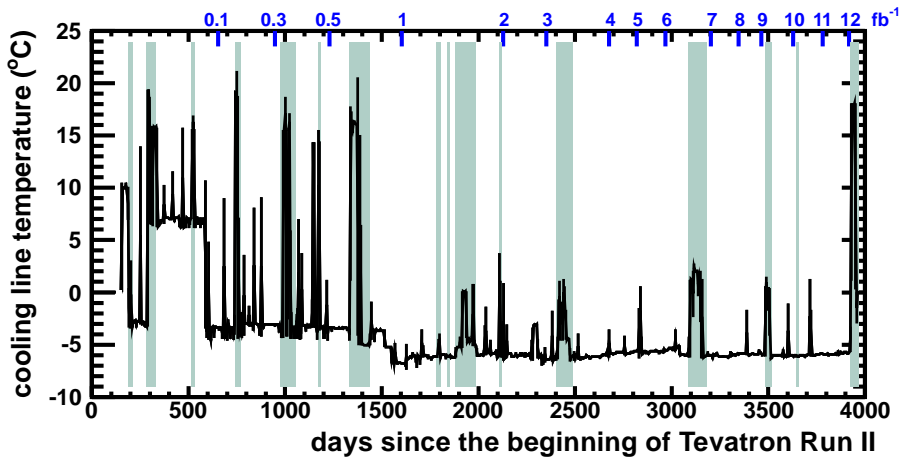}}
\label{fig:temps}
\caption{The average cooling line temperatures for L00 during Run II.
  Shutdowns are indicated with shaded bands.  Selected values of of
  integrated luminosity are indicated along the top of the plot at the
  time they were reached, but there is not a linear correspondence
  between time and luminosity.}
\end{figure}

During data-taking operations, the SVX sensors were sufficiently warm
that annealing and irradiation happened simultaneously throughout the
run. Annealing during shutdown periods was negligible compared with
operations for the SVX sensors for both bias current and depletion
voltage.  The L00 sensors were colder during operations.
Figure~\ref{fig:temps} shows the average measured cooling line
temperature for each day as a function of time during the run, with
shutdowns indicated with shaded bands.  After the first inverse
femtobarn of integrated luminosity, the coolant temperature was kept
below 5\degC even during shutdowns.  During shutdowns, the sensor
temperature was very close to the coolant temperature as the readout
chips were not powered and therefore the heat load on the system was
significantly reduced.  During operations, the sensor temperature is
estimated to be 2.5\degC warmer than the coolant, or -2.5\degC on
average.  A few small spikes in the daily average not associated with
a shutdown period are visible and result from excursions for a few
hours to warmer temperatures during short interruptions in coolant
circulation.

\subsection{Bias Current Evolution}
\label{sec:ibiasDiscussion}

The total change in bias current expected during the annealing period
can be predicted from Eq.~(\ref{eqn:ianneal}) and compared to the
observed values reported in Fig.~\ref{fig:currratio}.  The first three
terms in Eq.~(\ref{eqn:ianneal}) have time constants of 24 minutes,
2.7 hours and 1.0 days at 18\degC, or equivalently 18 hours, 5.0 days,
and 46 days at -5\degC, respectively.  Thus, the bias current
reduction due to annealing represented by the first two terms happened
during operations in a continuous fashion for L00. The contribution
from the third term is also anticipated to be negligble, considering
that less than 10\% of the total radiation dose was delivered in the
last 90 days of Run II.  Considering only the remaining three terms,
the measured ratio $I_{\mathrm{after}}/I_{\mathrm{before}}$ is
expected to be 68\% for 24 days of annealing at 18\degC, consistent
with the measured values for the L00 sensors.  For the SVX sensors,
only the last two terms are relevant for the annealing study, and 94\%
is expected for the ratio $I_{\mathrm{after}}/I_{\mathrm{before}}$,
again consistent with the measured values.
 
Similarly, we can compare the observed exponential decay of the bias
current with model expectations.  An experiment was simulated by calculating
expected measured currents with a modified version of
Eq.~(\ref{eqn:ianneal}) for the annealing time patterns of the actual
measurements, every 21 hours starting 1.5 days after warmup.  Using
the considerations of the previous paragraph, Eq.~(\ref{eqn:ianneal})
was modified by setting the amplitudes of the first three terms, $b_1,
b_2$ and $b_3$ equal to 0.

\begin{figure}
  \centering
  \includegraphics[width=0.8\textwidth]{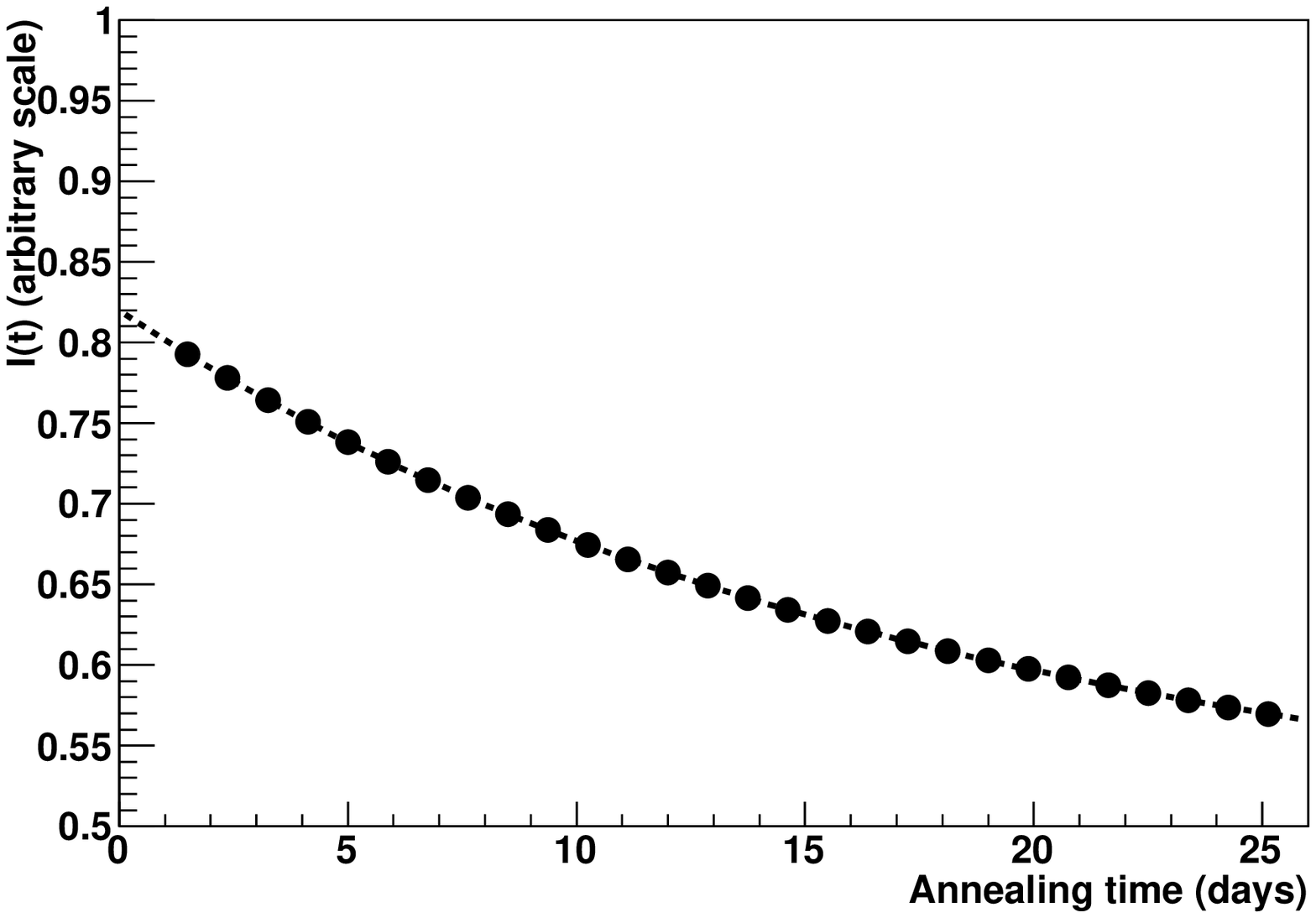}\\
  \includegraphics[width=0.8\textwidth]{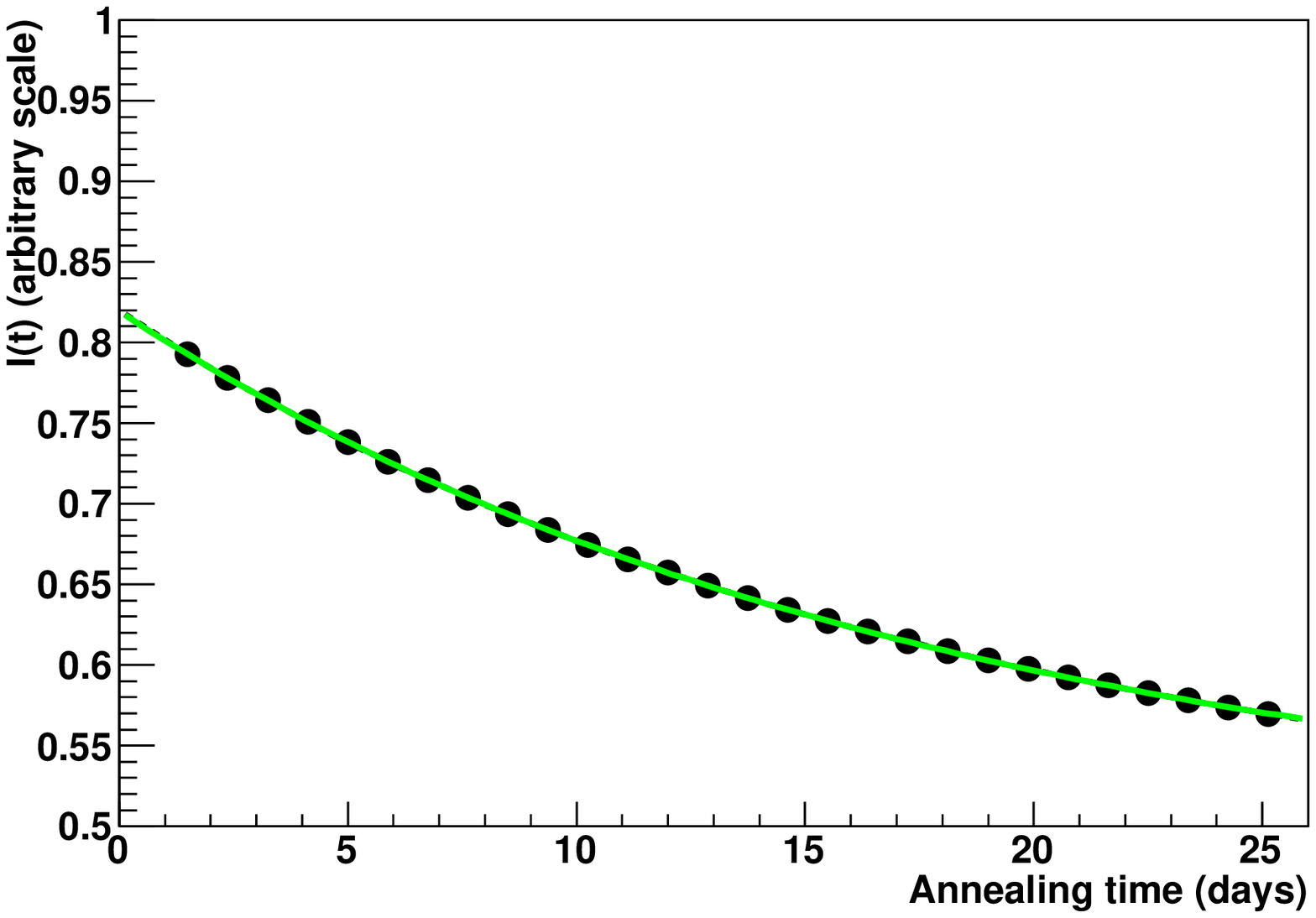}
\label{fig:model}
\caption{A simulated measurement of the decaying bias currents.  The
  upper plot shows with a dashed line the function used to generate
  the points, a modified version of Eq.~(\ref{eqn:ianneal}) with the
  amplitudes $b_{1,2,3}$ set to zero. The green solid line in the
  lower plot is the best fit result assuming
  Eq.~(\ref{eqn:iannealSimp}).}
\end{figure}

The upper plot of Fig.~\ref{fig:model} shows with a dashed line, the
expected measured bias currents as a function of annealing time from
the modified Eq.~(\ref{eqn:ianneal}).  In both plots, the circles
represent the expected measurements calculated by sampling the dashed
line every 21 hours starting at 1.5 days.  The solid line in the lower
plot is the best fit to the sampled currents using
Eq.~(\ref{eqn:iannealSimp}), giving a time constant of 17.58 days,
which is in good agreement with the measured value of
$17.88\pm0.36$(stat.)$\pm0.25$(syst.)~days.


\subsection{Depletion Voltage Evolution}

For the depletion voltage time constant, we expected $\tau_V =
3.6^{+2.2}_{-1.3}$ days.  Our measurement of $6.21 \pm 0.18$ days
exceeds this prediction by just over one standard deviation.  Due to
the complicated dependence on the various model parameters, it is
difficult to compare the measured after-before \vdepmap ratio of $(73
\pm 3)\%$ to the Hamburg model-prediction, which is somewhere between
50\% and 65\%, as mentioned in Sec.~\ref{sec:si_an}.  It does appear
that our measurement exceeds the Hamburg prediction to some
extent, which would be consistent with an assumption that annealing
happened to some extent during Run II.


%% file: summary.tex
\section{Summary}
\label{sec:sum}
After accumulating 12 fb$^{-1}$ of integrated luminosity, 
and being exposed to radiation doses up to 12 Mrad,
the CDF Run II silicon
detector was annealed at 18~\degC for 24 days.  The ratio of the bias
currents after the annealing study to that before the study
 is a measure of how much each
subsystem annealed during the run.  The overall change and evolution
of the bias currents, depletion voltages, and \vknee values of several
sensors during the annealing process were measured.

For L00, we observed a decrease of the bias current and depletion
voltage of the heavily irradiated sensors, with time constants
consistent with the Hamburg model expectations.  We observed no
indication of the reverse annealing on this time scale.

In contrast, the bias currents of SVX changed very little during the
annealing period.  This confirms that these sensors, with a
significantly warmer operating temperature than L00, annealed during
data taking. As the operating temperature of the SVX sensors was not
directly measured, and the prediction of their temperature from finite
element thermal models has large uncertainties, this is an important
confirmation of the annealing during the run, which extended the
lifetime of the detector.

\section{Acknowledgments}
The authors would like to thank CDF collborators Dr.~J.~Nett, 
Dr.~Y.Oksuzian, Dr.~I.~Redondo for the invaluable assistance in
collecting the data presented here and discussing analysis techiques.
The authors also thank
Dr.~A.~Junkes (Brown University) for
extensive discussions on radiation damage and annealing in silicon
detectors.
This work would not have been possible without strong
support from the CDF operations management and the spokespersons.  We
also thank the Fermilab staff and the technical staffs of the
participating institutions for their vital contributions.

This work was supported by the U.S. Department of Energy and National
Science Foundation; the Italian Istituto Nazionale di Fisica Nucleare;
the Ministry of Education, Culture, Sports, Science and Technology of
Japan; the Natural Sciences and Engineering Research Council of
Canada; the National Science Council of the Republic of China; the
Swiss National Science Foundation; the A.P. Sloan Foundation; the
Bundesministerium f\"ur Bildung und Forschung, Germany; the Korean
World Class University Program, the National Research Foundation of
Korea; the Science and Technology Facilities Council and the Royal
Society, UK; the Russian Foundation for Basic Research; the Ministerio
de Ciencia e Innovaci\'{o}n, and Programa Consolider-Ingenio 2010,
Spain; the Slovak R\&D Agency; the Academy of Finland; and the
Australian Research Council (ARC).

%% file: sh.tex
\section{Temperature corrections for dissipative heating}
\label{sec:sh}

The temperature dependence of the bias current for a fully depleted
reverse-biased diode is well understood~\cite{sze1985}. If the current
$I_0$ is measured at a known temperature $T_0$, then the current at
temperature $T$ is given by
\begin{linenomath}
\begin{equation}                                                               
  \frac{I(T)}{I_0}=\left(\frac{T}{T_0}\right)^2\exp\left[\frac{E_g}{2k_B}\left(     
      \frac{1}{T_0}-\frac{1}{T}\right)\right]                                        
\label{eqn:currscale}                                                          
\end{equation}
\end{linenomath}
where $k_B$ is Boltzmann's constant and $E_g=1.21\pm0.06$~eV is the
effective band gap energy~\cite{RD50ch}.

The temperature of a biased sensor increases as a result of resistive
power dissipation in the sensor, hereafter referred to as
``self-heating''.  For the \iv scans of the annealing period, this
temperature increase was as large as 3\degC for the L00 narrow sensors
at the largest voltages.  Such a temperature shift can result in bias
current deviations from nominal 18.2\degC values by as much as 30\%.
These temperature variations must be removed from a set of measured currents.
before information about the annealing processes can be extracted.

The measured currents can be corrected to a common temperature using
Eq.~(\ref{eqn:currscale}) if the sensor temperature at the time of the
measurement is known.  The sensor temperature is not measured
directly, however it can be calculated from the measured current $I$
at a given bias voltage $V$, assuming that the temperature increases
linearly with the power dissipated by the ladder:
\begin{linenomath}
\begin{equation}
T=18.2\degC + \kappa IV
\label{eqn:kappa}
\end{equation}
\end{linenomath}
where $\kappa$ is a proportionality constant.

Using values of $\kappa$ determined with a dedicated measurement, a
temperature correction was applied to the measured currents before
extracting the knee voltage, as shown in Fig.~\ref{fig:ex_scan27_cent}, and
the time constant of the current decay, shown in
Fig.~\ref{fig:narrowCurr}.

\subsection{Determination of $\kappa$ for each sensor}

The value of $\kappa$ for each sensor was determined by combining data
from a dedicated reference current measurement with the last set of
\iv scans taken during the annealing period.  A collection of
reference current measurements were taken at the operating voltage
$V_\mathrm{op}$ and at half of the operating voltage $0.5\times
V_\mathrm{op}$ after the 24-day annealing period.  For these
measurements, the readout chips were not powered and the measured
cooling line temperatures were between -0.5 and 1.1\degC for the
system.  Only one sensor was biased at a time to minimize any
potential heat load.

The expected bias current $I_{\mathrm{ref}}^{\, \prime}$ at the
annealing temperature $T_a=18.2$\degC, in the absence of self heating,
can be calculated from the measured reference currents
$I_{\mathrm{ref}}$
\begin{linenomath}
\begin{equation}
I_{\mathrm{ref}}^{\, \prime}=I_{\mathrm{ref}}\left(\frac{T_a}{T_{\mathrm{ref}}}\right)^2\exp\left[\frac{E_g}{2k_B}\left(     
      \frac{1}{T_{\mathrm{ref}}}-\frac{1}{T_a}\right)\right]    
\end{equation}
\end{linenomath}
where $T_{\mathrm{ref}}$ is the temperature of the sensor during the
measurement of the reference currents.

The last set of \iv scans taken during the annealing period contain
bias currents measured in this voltage range.  Self heating increases
the temperature of the sensor to a value $T_{\mathrm{hot}}=18.2\degC +
\kappa IV$, and the measured current $I_{\mathrm{meas}}$ can be
corrected to the equivalent current $I_{\mathrm{corr}}$ at the
annealing temperature $T_a=18.2$\degC by 
\begin{linenomath}
\begin{equation}
I_{\mathrm{corr}}=I_{\mathrm{meas}}\left(\frac{T_a}{T_{\mathrm{hot}}}\right)^2\exp\left[\frac{E_g}{2k_B}\left(     
      \frac{1}{T_{\mathrm{hot}}}-\frac{1}{T_a}\right)\right] \ .
\end{equation}
\end{linenomath}
Because there are two reference current measurements at two different
bias voltages for each sensor, unique values of $T_{\mathrm{ref}}$ and
$\kappa$ for each sensor are determined by requiring that
$I_{\mathrm{corr}}=I_{\mathrm{ref}}^{\, \prime}$ for both
measurements.  This initial result is then improved upon iteratively
by correcting the reference currents for a small amount of self
heating, using the values of $\kappa$ and $T_{\mathrm{ref}}$ from the
previous iteration in Eqs.~(\ref{eqn:kappa})
and~(\ref{eqn:currscale}).  For all ladders, the values of $\kappa$
and $T_{\mathrm{ref}}$ converge in fewer than four iterations.  The
difference between the initial and final values of $\kappa$ is taken
as a systematic uncertainty.

\begin{figure}
  \centering
    \includegraphics[scale=0.6]{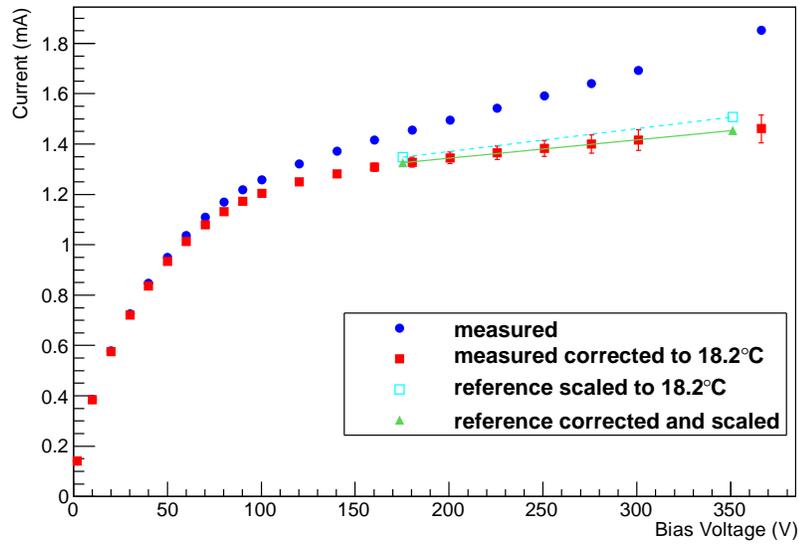} 
    \caption{The measured bias current as a function of bias voltage
      for a typical L00 narrow ladder is plotted with blue circles.
      The equivalent current at 18.2~\degC\ is shown with solid red
      squares.  The open cyan squares show the reference measurement
      scaled to 18.2~\degC\ and the green triangles the reference
      measurement first corrected for self-heating and then scaled to
      18.2\degC.  The values $\kappa=3.925$~K/W and
      $T_\mathrm{ref}=-0.4$\degC give the best agreement between the
      red squares and the solid green line for this ladder.  }
    \label{fig:tempcorr}
\end{figure}

Figure~\ref{fig:tempcorr} illustrates this process for a typical
narrow L00 ladder. The measured currents $I_{\mathrm{meas}}$ from the
\iv scan are shown as solid circles, and the corrected currents
$I_{\mathrm{corr}}$ as solid squares. The values of $\kappa$ and
$T_{\mathrm{ref}}$ from the first iteration are those for which the
solid squares agree with the dashed line, while the final values give
agreement with the solid line connecting the corrected reference
currents shown with solid triangles.  If the bias voltage value of the
\iv scan points do not exactly match those of the reference
measurement, as in the example illustrated, then the two \iv scan
points closest to the reference scan measurements are chosen, and a
linear interpolation is used to obtain measured reference currents for
those voltages.

\begin{figure}
  \centering
  \includegraphics[scale=0.6]{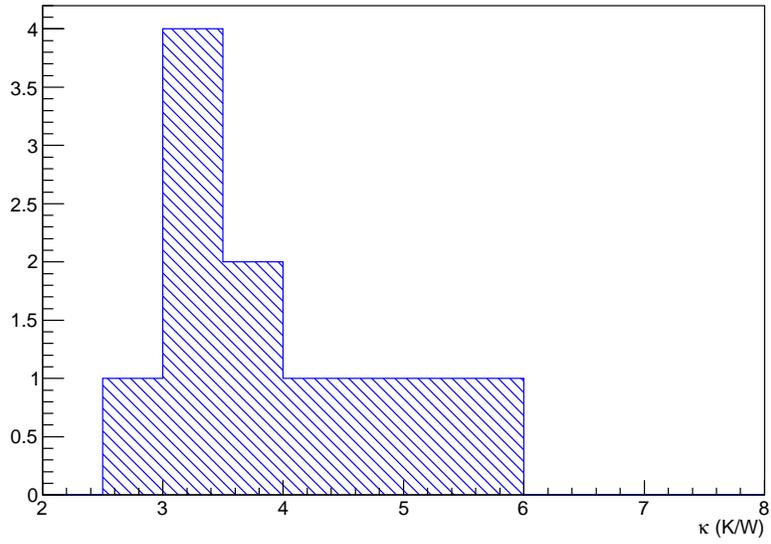}
  \caption{The distribution of $\kappa$ for the L00 narrow ladders.}
  \label{fig:kappa}
\end{figure}
\begin{figure}
  \centering
  \includegraphics[scale=0.6]{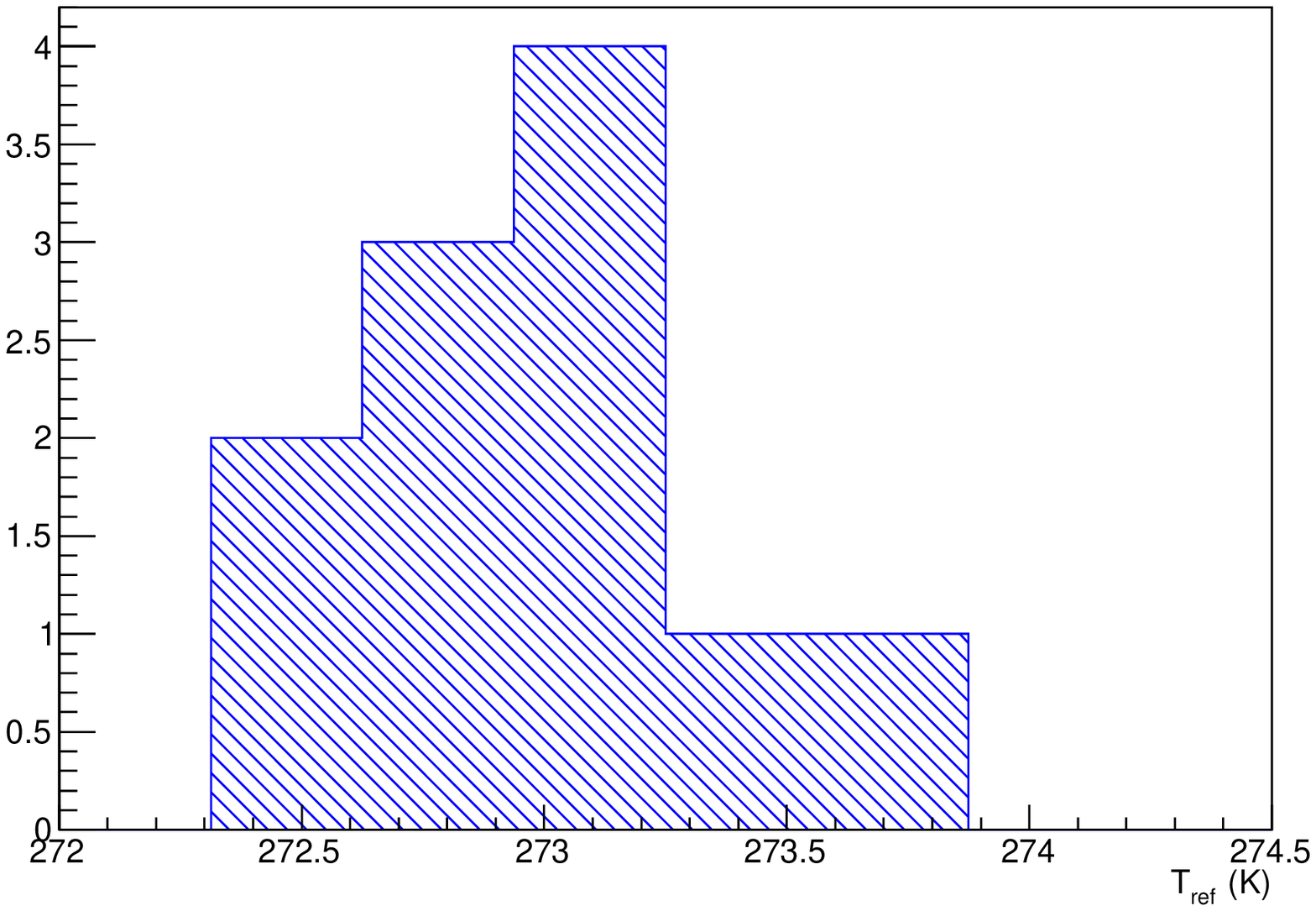}
  \caption{The distribution of $T_\mathrm{ref}$ for the L00 narrow
    ladders.}
  \label{fig:trdist}
\end{figure}

The distributions of $\kappa$ and $T_{\mathrm{ref}}$ are shown in
Figs.~\ref{fig:kappa} and \ref{fig:trdist}.  These values are expected
to vary slightly with the ladder location and thermal connection to
the cooling lines.  $T_{\mathrm{ref}}$ and $\kappa$ have common
systematic errors from the uncertainties on the absolute annealing
temperature $T_{\mathrm{warm}}=18.2\pm0.5$~\degC and the effective gap
energy $E_{g}=1.21\pm0.06$~eV~\cite{RD50ch}.  For $\kappa$, these two
uncertainties combine to give an overall uncertainty of $0.05$~K/W.
The time delay between the last warm \iv scan and the reference
measurement is different for each ladder, varing from 3 to 24 hours.
The current at fixed temperature and voltage decreases due to
annealing during this time, but the change is observed to be less than
0.5\% for the longest time period and the resulting shifts in
$T_{\mathrm{ref}}$ and $\kappa$ are negligible.  An important
verification of this self-heating correction method comes from the
agreement of the best fit values of $T_{\mathrm{ref}}$ with the
expectation $T_{\mathrm{ref}}=273.0\pm1.0\pm1.0$~K derived from the
measured cooling line temperatures.